\documentclass[aps,prb,showpacs,twocolumn,floats,epsfig,pdflatex]{revtex4-1}
\usepackage{epsfig}
\usepackage{epstopdf}
\usepackage{amsmath}
\usepackage{amsfonts}
\usepackage{graphicx}
\usepackage{amssymb}
\usepackage{amsbsy}

\newcommand{\cs}{$\clubsuit$ \bf}

\begin{document}

\title {Triplet Superfluidity on a triangular ladder with dipolar fermions}

\author{Bradraj Pandey$^{(1)}$,  and Swapan K. Pati$^{(1)}$}
\affiliation{$^{(1)}$ Jawaharlal Nehru Centre for Advanced Scientific Research, Jakkur P.O., Bangalore-560064, India.
 }


\date{\today}

\begin{abstract}
Motivated by recent experimental progress in the field of dipolar-Fermi gases, 
we investigate the quantum phases of dipolar fermions, 
on a triangular ladder at half filling. 
Using density matrix renormalization group method, in presence of onsite 
repulsion and intersite attractive interaction, we find exotic 
spin-triplet superfluid phase in addition to the usual spin-density and 
charge-density waves. 
We examine the  stability of spin-triplet superfluid phase by varying hopping along 
the rungs of the triangle. Possibility of fermionic supersolidity has also been discussed, 
by considering three-body interaction in the Hamiltonian. 
We also study the effect of spin-dependent hopping on the stability of 
spin-triplet superfluid phase.
\end{abstract}


\maketitle

\section{Introduction}
Recent experimental advancements in the field of dipolar 
Fermi gases have given opportunity to explore the quantum phases 
of strongly correlated fermionic 
systems with long-range interactions
\cite{amicheli,tlahaye}.
The dipolar Fermi gas of $^{161}Dy$ \cite{Mingwu} and fermionic 
polar molecules, $^{40}K^{87}Rb$ \cite{chotia}, 
$^{23}Na^{40}K$\cite{chwu}, with large dipole moments have
experimentally been realized in optical lattices.
It has been found that the external electric and microwave
fields on optical lattices can control quantum many body 
interactions parameters of dipolar systems and polar molecules
\cite{mmarin,k-k,lsan,imour}. It has been argued that the
long range and anisotropic characters of the dipolar interactions,
in fact, can provide various types of exotic phases like, 
charge-density wave (CDW; even though the density modulation is 
produced by charge neutral atoms or molecules, it is called CDW
in the literature)\cite{mmparish,KMikelsons,YYamaguchi},
spin density wave (SDW; spin order for pseudo-spin-1/2 of dipolar fermions,
shown in schematic of Fig.2(a)) \cite{WenMin,SGBhongale}, liquid-crystal\cite{CLin,JQuintanilla}, 
conventional and unconventional fermionic superfluids 
\cite{LYou,MABaranov,TShi,GMBruun,NRCooper,
BoYan}, to name a few.

Finding phases, like, triplet superfluidity and triplet superconductivity
are always very challenging and interesting too as these exotic phases
have connection to a number of topological phases and quantum computation.
Interestingly, at low temperature, liquid $^3He$ forms 
fermionic superfluids, where $^3He$ atoms (or quasi particles) 
form pairs with p-wave symmetry in spin triplet 
state\cite{DVollhardt,Anthony}.
Chromium based quasi-one dimensional superconductors
\cite{jinke,xwu} and strontium based oxide, $Sr_2RuO_4$, 
are considered to be good candidates for triplet 
pairing\cite{APMackenzie,YMaeno}. 
 
Interestingly, ultra cold dipolar systems, offer 
intriguing possibilities to explore
unconventional pairing mechanisms of the condensed-matter system.
For single component fermions, a dominant $p_z$-wave superfluidity
has been proposed\cite{LYou,MABaranov}. For two components 
fermions, it has been shown that there is possibility of 
formation of both singlet and triplet superfluidity\cite{Congjun,
RQi,KSun}, as both singlet and triplet pairing are allowed 
in such systems. In two-dimensional dipolar fermionic system,
 where dipoles are aligned with external electric field, it
has been shown that p-wave superfluidity can be realized 
by varying anisotropy and geometry of the system\cite{GMBruun}. 
Unconventional spin-density waves\cite{SGBhongale} 
and  bond-order solids\cite{sgl}
have also been shown for the two-dimensional dipolar systems.

On the otherhand, more exotic phases, like, supersolid phase,
has been proposed for dipolar Fermi gas in a cubic optical 
lattice system\cite{tian}. Interestingly, in this, it has been
shown that a $p$-wave superfluid is formed due to attractive 
interaction along the z-direction, and charge-density wave 
in the XY-plane due to electronic repulsions and together with
the intermediate values of dipolar interactions. For a two 
dimensional dipolar Fermi gas, coexistence of density-wave and 
p-wave superfluidity has been shown\cite{zwu,lhe}. 
In a recent experimental study on ultra-cold three dimensional
optical lattice systems, effect of multi body interaction has 
been demonstrated\cite{new,swill}.
Furthermore, in a few numerical studies, it was shown that 
dominant three body Coulombic interactions can give rise to a host 
interesting phases, like supersolid and bond-order phases
\cite{zhenkai,xuefeng,bcapogr,tmishra}. Interestingly, for 
polar molecules in optical lattice, the realization of 
three-body interactions using microwave field have been proposed
\cite{hpb,kps,lbonnes} and since then there have been various 
theoretical studies of microscopic models with three-body 
interactions\cite{rdmur,nrcoo,dspet,awojs,jkpac}. These studies
have shown that, with three body Coulombic interactions, the
ground state can be quite exotic displaying quantum phases like,
topological phases, spin liquids, fractional quantum Hall states
etc. 

Quasi one dimensional systems are quite unique. Due to 
strong quantum fluctuations, the true long range order is 
not possible for continuous symmetry breaking 
phases\cite{macazilla}. In a one dimensional optical lattice, 
bosonization study has shown triplet superfluid (TSF) phase
for dipolar fermions\cite{tnde}. TSF phase is also found in 
two coupled one dimensional systems for quadrupolar 
Fermi gas\cite{wenmin}. Interestingly, mixture of triplet 
and singlet superfluidity has also been shown in a quasi-one 
dimensional system with two component fermions\cite{suchino}.
A recent DMRG study\cite{hmosad} study has also found the TSF phase in a 
one-dimensional dipolar Fermi gas. 
In presence of attractive head to tail arrangement of dipolar 
interactions, the one and two dimensional 
dipolar fermions become unstable and they undergo either 
collapse or phase separation. To overcome these difficulties, 
bilayer system has been proposed, where dipoles are aligned 
perpendicular to the layers, giving more stable paired 
phases\cite{acpotter,dawwei}.  
  
In this article, we consider dipolar fermions in a 
triangular ladder system at half-filling. We study the
stability of various exotic phases, like, spin-density wave, 
charge density wave and triplet-superfluid phases.
In the ladder, the dipolar fermions are considered to be 
polarized along the rungs of the triangles (as shown 
in schematic of Fig.1). 
The strength and direction of polarization can be 
controlled by external electric field or by varying 
distance between lattice sites. Due to alignment of 
dipolar fermions along the rungs, attractive interaction 
is generated on alternative rungs (odd rungs). It is also possible 
to generate repulsive interaction in each of the chains and 
diagonal rungs of triangle, by alignment of dipoles. 
In the presence of attractive dipolar interaction and on-site 
Hubbard repulsion, a stable TSF phase gets generated.
We have checked the stability of the TSF phase thoroughly, 
by tuning in the inter-chain hopping strength and the 
repulsive interaction parameters. Additionally, We have 
also examined the effect of spin-dependent interchain 
hopping on the stability of the TSF phase.
Interestingly, due to 
triangular geometry, three-body interactions can 
also play important role in identifying new quantum 
phase, like, fermionic super-solid phase of dipolar 
fermions\cite{tmishra}. 

The remaining part of the article is organized 
as follows. 
In sec.II we have discussed the model Hamiltonian and
the method used to solve it. Subsequently, we have discussed
the results obtained from DMRG calculations. This is divided 
into four subsections, where in each subsection the details of
phase and phase transition is discussed. In last section,
we have summarized all our results.

\section{The Model}
\begin{figure}[h]
\rotatebox{0}{\includegraphics*[width=\linewidth]{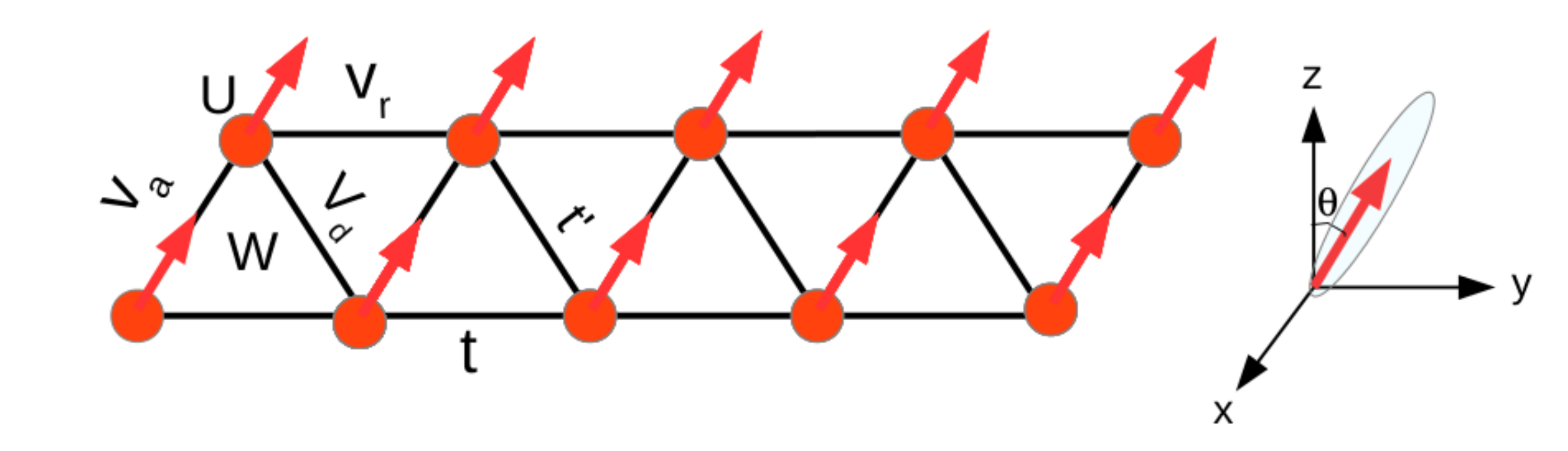}}
\caption{Schematic of the triangular ladder with dipolar 
fermions (arrows indicate the directions of polarization of fermionic dipoles). 
There is onsite interaction $U$, attractive interaction $V_a$, repulsive interactions $V_r$, and $V_d$. Three body interaction term is given
 as $W$ and the hopping along the legs and rungs are represented as $t$ and $t'$}.
\end{figure}

We consider two-component (pseudo-spin-1/2) dipolar fermions on a two-leg triangular ladder at half-filling.
The effective Hamiltonian of the  system can be written as,
\begin{eqnarray*} 
H= -\sum_{\sigma, i} \left(t c^{\dagger}_{\sigma,i} c_{\sigma,i+2} + 
t' c^{\dagger}_{\sigma,i} c_{\sigma,i+1} + H.c\right) + \\
U\sum_i \hat{n}_{i,\uparrow} \hat{n}_{i,\downarrow} 
+ \sum_{\langle i \ne j \rangle} V(i,j)\tilde{n}_{i} \tilde{n}_{j} - W \sum_i \tilde n_i \tilde n_{i+1} \tilde n_{i+2}
\end{eqnarray*}
\noindent where  $c_{\sigma,i}$ is annihilation operator with spin $\sigma= \uparrow,\downarrow $
at site $i$. Here $\uparrow$ and $\downarrow$ states refer to two hyperfine states of diploar atoms or molecules.
$\tilde{n}=(n-\langle n \rangle)$ where n is the number operator and $\langle n \rangle=1$ .
 $t$ and $t'$ are  the hopping terms and
$U$ is the onsite interaction term between the fermion with opposite spins; $V(i,j)$ 
is the two-body nearest-neighbour intersite interaction term.
The last term in the Hamiltonian, $W$, 
represents attractive three body interactions between the fermions, 
which act on the fermions belonging to the same triangle (as shown in the Fig.1).
The two-body interaction term depend on direction and distance between the dipoles.
When the two dipoles are parallel to each other, the interaction
becomes repulsive, while when they align to each other along the rungs, 
interaction become attractive. The most dominating interactions 
arise from the nearest-neighbour terms\cite{tref,xiao}, and also in optical 
lattice by adjusting the distance between sites, one can make other subdominating 
interactions quite smaller\cite{xiao}. Thus, we restrict ourself to only nearest-neighbour terms of 
$V(i,j)$ in the Hamiltonian\cite{tmishra}. 
The two body nearest-neighbour term, $V(i,j)$, can be described as
\begin{displaymath}
V(i,j) = \left\{ \begin{array}{ll}
V_{r}  & \textrm{Intersite repulsive term on each chain.}\\
V_d & \textrm{Intersite repulsive term for even rungs.} \\
-V_a & \textrm{Intersite attractive term for odd rungs.}
\end{array} \right.
\end{displaymath}
Since the dipolar interaction depends on angle and distance between the dipoles,
it allows tuning of magnitude and sign of these interaction parameters to a wide range
to explore rich quantum many-body phases. 
The dipolar interactions can be tuned by external electric field or changing the distance between sites. 
The above Hamiltonian preserves $U(1)$  and $SU(2)$ symmetry,
which is related to conservation of total charge and 
spin degrees of freedom. Note that, for nonzero 
next nearest neighbor terms, $t$ and $W$, the Hamiltonian
does not have particle-hole symmetry.
 
To solve the above Hamiltonian and to find  quantum phases in the
parameter space, we have used density-matrix renormalization 
group (DMRG)\cite{White,Schollwock} method.
We have used open boundary conditions and vary the DMRG 
cut-off (max = m) from $300$ to $600$, for consistency in results.
Most of the results presented in the article are obtained using max=$520$, 
unless otherwise stated.
To calculate the error, we have checked the truncation error,
 $e=1-\sum_i\rho_i$, where $\rho_i$ is the eigenvalues corresponding to the
reduced density matrix. We found that depending upon the interaction parameters
and system size, truncation error $e$ varies from $10^{-5}$ to $10^{-6}$. 
We have verified energy and excitations for some parameters 
with those from exact diagonalization for smaller system sizes. 
To characterize different phases, namely $SDW$, $TSF$, and $CDW$ 
phases, we have calculated corresponding correlation 
functions and also spin and charge density profiles.
For showing plots of correlation functions, unless stated
explicitly, we have considered system size $L=128$.
To determine phase boundary between different
phases and to minimize the finite size effect, we have done
finite-size scaling of order-parameters, of the system with size
(L) up to $160$.

\section{Results}
\subsection{SDW to TSF to CDW transition}
\begin{figure}[h]
\rotatebox{0}{\includegraphics*[width=\linewidth]{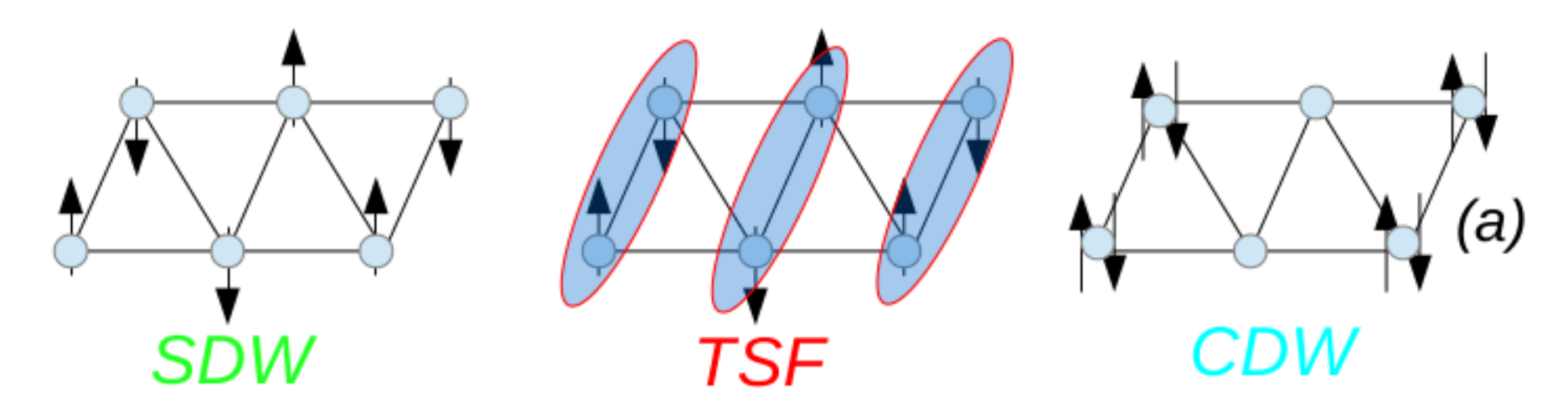}}
\rotatebox{0}{\includegraphics*[width=\linewidth]{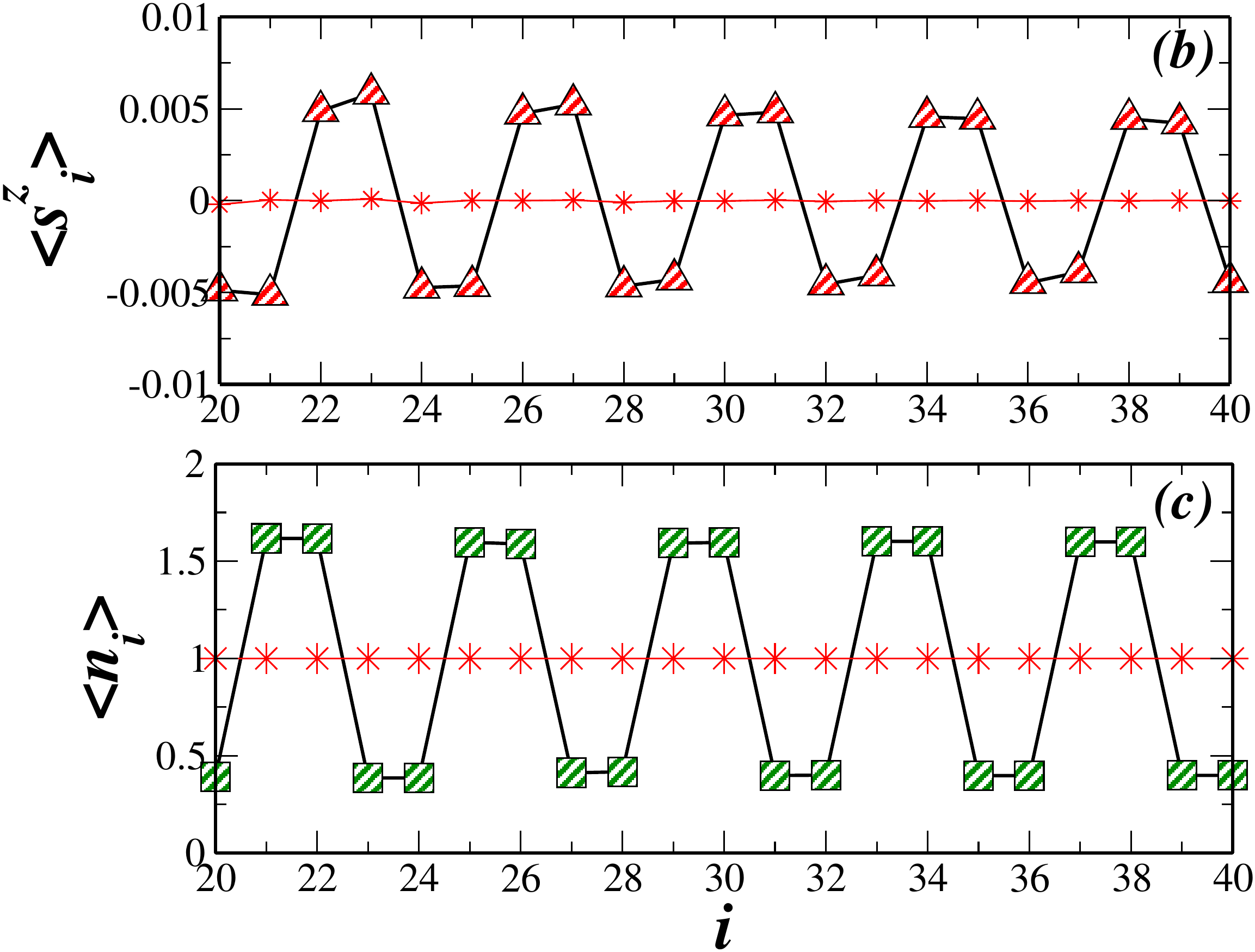}}
\caption{
(a) Schematic of the SDW, TSF and CDW phases on a triangular lattice 
(here arrows indicate electronic spins of Fermions).
(b) Plot of spin-density $\langle s^z_i \rangle$ with site index $i$, 
for $V_a=1.6$ (triangle) and $V_a=2.5$ (star). (c) Plot of charge density
 $\langle n_i \rangle$ for $V_a= 2.4$ (star) and $V_a=3.2$ (square).} 
\end{figure}

We first consider a simple case, where $t'=0$, the intersite repulsive dipolar term, 
$V_r=0$, $V_d=0$ and three body term, $W=0$.
Due to long range of dipolar interactions, two 
chains of traingular ladder can couple through attractive dipolar interaction, $V_{a}$, 
even though the tunneling between the chains remain zero\cite{tnde}.
For finding TSF phase, we take onsite Hubbard interaction $U=2$, and vary
the attractive interaction, $V_a$ ($0$ to $4$), along the rungs (odd rungs). 
For $U=2$ and lower vales of $V_{a}$, we find
that to minimize repulsive onsite interaction, fermions stay 
put in each site and form spin density wave, 
$|\uparrow, \uparrow,\downarrow,\downarrow,\uparrow,\uparrow,
\downarrow,\downarrow,\uparrow, \uparrow... \rangle$ 
(as shown in schematic of Fig.2(a)). In order to show
spin density profile of the system, in Fig.2(b), 
we have plotted  spin-density $\langle s^z_i \rangle $ 
of system, with site index, $i$.
With increase in attractive interaction, $V_{a}$, the 
fermions form intersite pairs along the rungs of the ladder,
where the electronic spins form triplet symmetry ($|s^z=0\rangle= 
|\uparrow \downarrow \rangle + |\downarrow \uparrow \rangle $)
\cite{aromano}. 
This phase remains so for moderate values of $V_a$.
For large value of attractive interaction, fermions with up 
and down spin prefer to sit together and 
form CDW-phase, where the state appears like, 
$|\uparrow \downarrow, \uparrow \downarrow, 0,0,
\uparrow \downarrow,\uparrow \downarrow,0,0 ... \rangle$ 
(as shown in schematic of Fig2(a)). 
To show this, in Fig.2(c), we have plotted charge 
density profile of fermions, $\langle n_i\rangle$, 
with site index $i$. Interestingly, this CDW-phase 
appears even without any intersite-repulsive terms.
Thus is precisely due to the triangular geometry and 
the attractive interaction along 
leg-direction\cite{fmarche}.
However, in strictly one dimensional case, for large 
values of attractive interaction, the system goes to either 
phase-separated phase or it collapses\cite{hmosad}.

In order to characterize SDW, TSF and CDW phases and their 
boundaries, we vary $V_a$ with fixed value of $U= 2$, and 
we look into the behavior of corresponding correlation 
functions. For SDW phase, we have calculated correlation 
function, $S(r)=\langle s_i^z s_{i+r}^z \rangle $, where $r$ (even distances) is
the distance from the middle site of the ladder to the one end of the ladder.
We found that with increase in $r$, fluctuations appears in the correlation function (Appendix Fig.15).
To reduce these fluctuations, we have calculated average correlation function,   
$S(r)= 1/N(r)\sum_r \left|\langle s_i^zs_{i+r}^z\rangle \right|$.
Here, we have summed over the all correlations, which are separated
by the same distance $r$ from sites $i$ and divided by the numbers $N(r)$
of such same distances correlations \cite{mguerro}    
. As shown in Fig.3(a), for lower values of $V_a$, the 
correlation function, $S(r)$, decays algebraically, 
while for larger values of $V_a \gtrsim 2.0$, it decays exponentially.
\begin{figure}[h]
\rotatebox{0}{\includegraphics*[width=\linewidth]{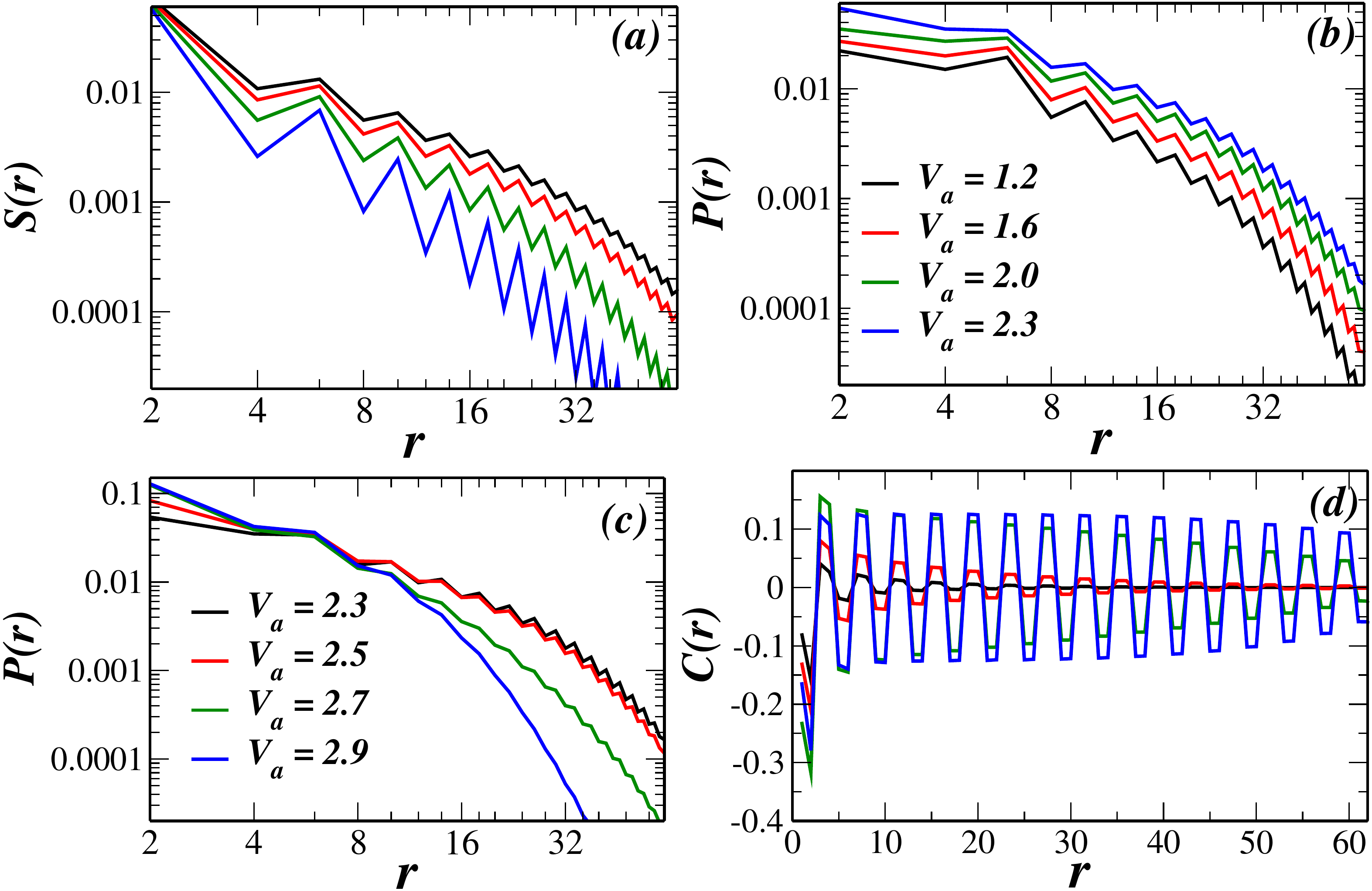}}
\caption{
(a) Plot of correlation function $S(r)$, (b) correlation function $P(r)$, for $U=2$ and varying $V_a < 2.3$.
(c) Plot of correlation function $P(r)$, (d) correlation function $C(r)$, 
for $U=2$ and varying $V_a$ (2.3 to 2.9).
}
\label{fig3}
\end{figure}
\begin{figure}[h]
\rotatebox{0}{\includegraphics*[width=\linewidth]{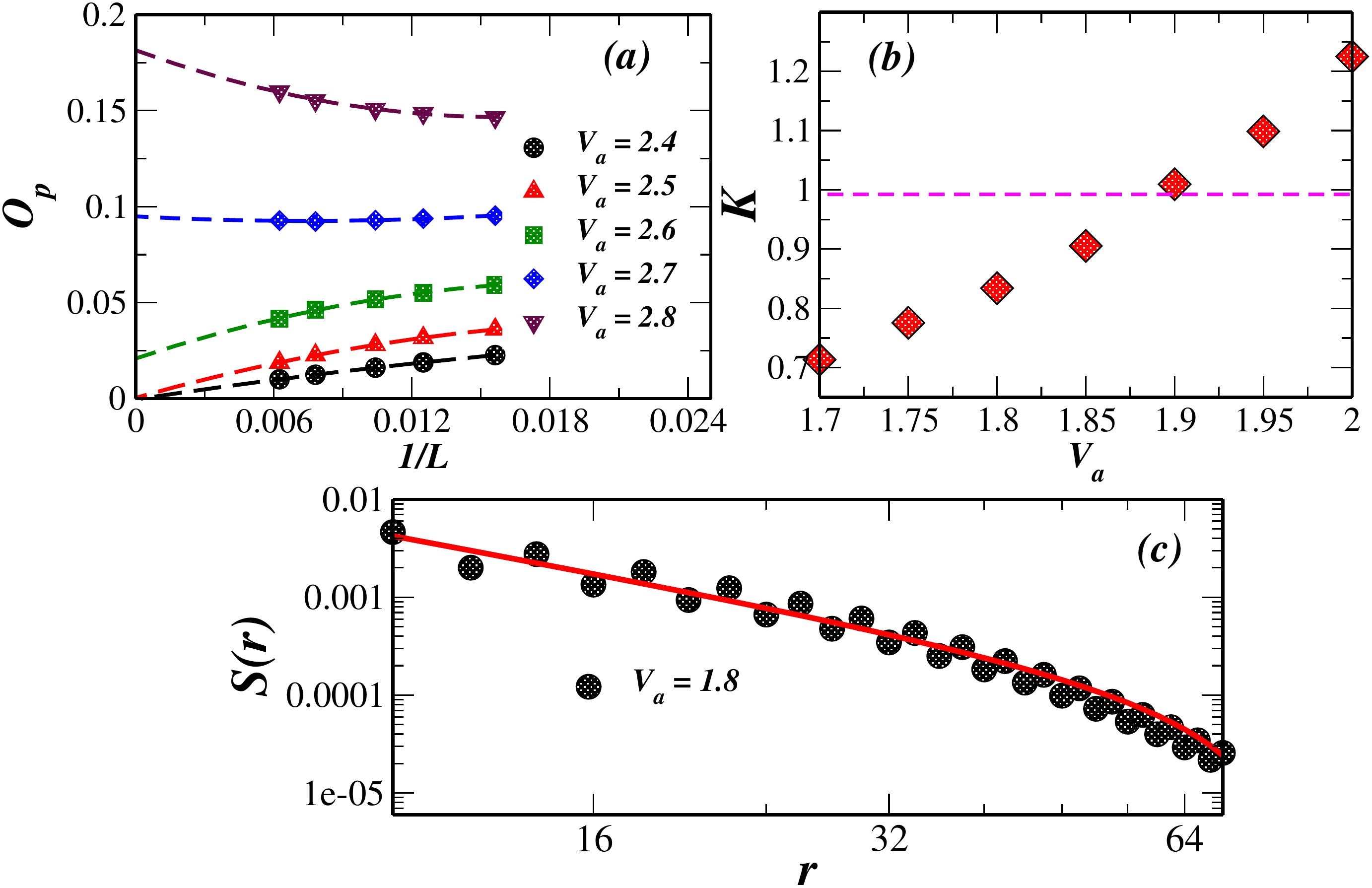}}
\rotatebox{0}{\includegraphics*[width=\linewidth]{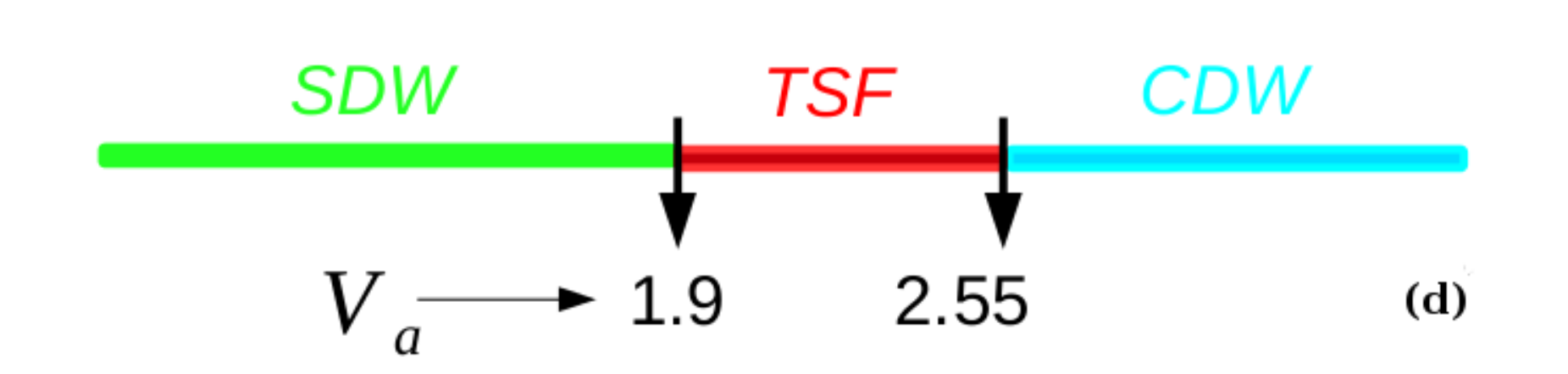}}
\caption{Finite-size scaling of (a) order parameter $O_p$ (b) exponent $K$ of the correlation function
$S(r)$, at $U=2$ and different values of $V_a$. 
(c)Power law fitting of $S(r)$ at $V_a = 1.6$, on a log-log scale for system size $L=128$. 
(d) phase diagram for fixed value of $U=2$ with varying $V_a$.
}
\label{fig4}
\end{figure}

With increase in attractive attraction along the rungs of 
the triangle, interchain fermions form bound pairs along 
the rung, giving rise to interchain spin-triplet superfluid 
phase, which is quite interesting. 
In general, the TSF phase can be characterized by 
pair correlation function\cite{mtez,giamar,fiemini}
$P(r)= \langle \Delta_l^+ \Delta_{l+r}\rangle$,
\noindent where $\Delta^{\dagger}(l)= \left(c_{i,\uparrow}^{\dagger}
c_{i+1,\downarrow}^{\dagger}+c_{i,\downarrow}^{\dagger}
c_{i+1,\uparrow}^{\dagger}\right)$, creates a fermionic pair in spin triplet state
on a rung (labeled $l$) and 
$r$ (even distance) is the distance from the rung $l$ (near to the 
center of the triangular ladder).
This correlation function $P(r)$, is also called $p_z$ wave like superfluid 
correlation function, because of spin triplet pairing along z direction.
 For $P(r)$ also, fluctuations appear with increase in $r$.
To smooth out these fluctuations, we have calculated average correlation function, 
$P(r)= 1/N(r)\sum_r |\langle \Delta_i^+ \Delta_{i+r}\rangle|$, 
where we have summed over the correlations which are separated 
by same distances $r$ from rung $l$, divided by the numbers, $N(r)$, with such same 
distances correlations. 

To characterize the phase boundary accurately between $SDW$ and $TSF$ phases, we have calculated the  
exponent of the correlation function, $S(r)$.  The exponent, $K$, can be obtained by fitting the
 correlation function with algebraic decay function of the form, 
$S(r) \sim cos(2k_Fr)(1/r)^{1+K}$ (as shown in Fig. 4(c))\cite{arra,troyer}.
To get rid of short range correlation functions and the finite size effects, 
we have fitted the correlation function, $S(r)$, from distance r =10 to 70, for system size length L=160.
We find that the correlation function, $S(r)$, fits very well in the SDW phase, however, near the 
phase boundary close to the $TSF$ phases, the fitting error increases. 
From Luttinger liquid theory, for $K<1$, the $SDW$ phase dominates, while for $K>1$, the $TSF$ phase 
dominates \cite{karlo,torsten,arra,troyer}.  The transtion point for $SDW$ to $TSF$ phase is expected 
to be at $K=1$. As shown in the Fig. 4(b), at $V_a=1.9\pm0.06$, the exponent $K$ of the correlation function $S(r)$ 
takes the value $K\sim 1$, which signifies the transition from $SDW$ phase to $TSF$ phase.

To characterize CDW-phase, we have calculated correlation 
function, $C(r)=\langle (n(i)-\bar{n})(n(j)-\bar{n})\rangle$.
where $r$ is the distance from middle site of the ladder to other on one side of the ladder. 
As shown in Fig.3(d), the correlation function, $C(r)$, 
for $V_{a} > 2.5$ has nearly long range order, 
while $P(r)$ decays exponentially (as shown in Fig3.(c)). 
Thus, for $V_{a}> 2.5 $, the system is in the 
CDW phase. To calculate the phase boundary between
$TSF$ and $CDW$ phase, we have done finite size scaling of 
order-parameter, $O_p= (1/L)\sum_{r=1}^L|C(r)|$.
In the density wave phase order-parameter $O_p$, takes non-zero values
in the thermodynamic limit\cite{zhang}. To obtain the thermodynamic value of
$O_p$, we have done finite-size scaling for
systems with length $L$ up to $160$, by fitting the finite-size
$O_p$ values with a function, $O_p + O_1/L + O_2/L^2$.
As shown in the Fig.4(a), $TSF$ to $CDW$ transition occurs at $V_a=2.55\pm0.05$ 
as $O_p$ takes finite non-zero values for $V_a=2.55\pm0.05$.

 As shown in schematic of Fig.4(d), for fixed values 
of onsite interaction, $U=2$ and by varying $V_a$, we 
found SDW phase for $V_a \lesssim 1.9$, TSF phase for 
$1.9\lesssim V_a \lesssim 2.55$ and CDW-phase for $V_a \gtrsim 2.55$.

\subsection{Effect of Onsite Repulsive Interaction}
\begin{figure}[h]
\rotatebox{0}{\includegraphics*[width=\linewidth]{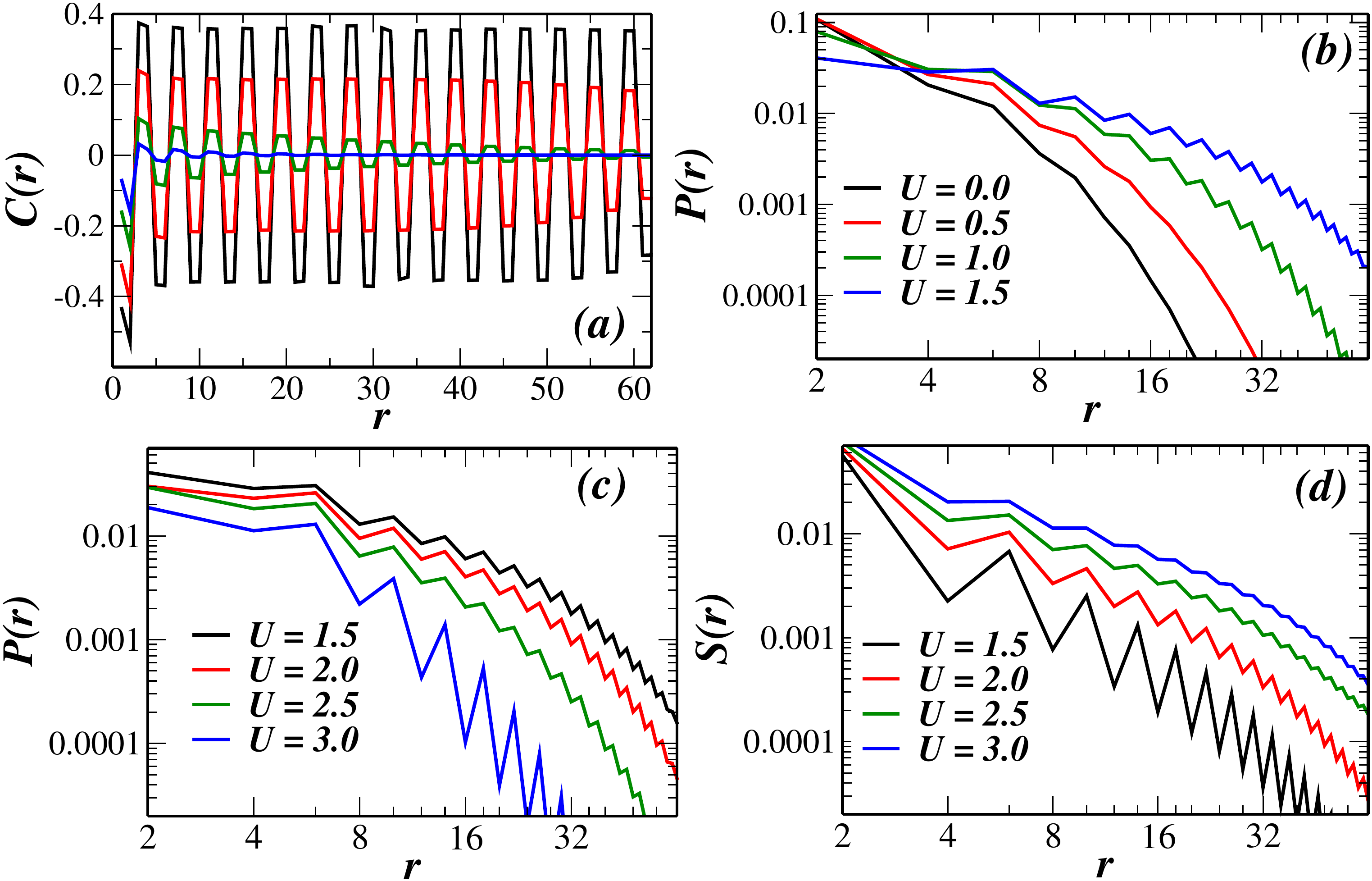}}
\caption{
(a) Plot of correlation function $C(r)$, (b) correlation function $P(r)$, for $V_a =1.8$ and varying $U < 1.5$.
(c) Plot of correlation function $P(r)$, (d) correlation function $C(r)$, for $V_a=1.8$ and varying $U$ (1.5 to 3.0).
}
\label{fig5}
\end{figure}

To find the role of onsite interaction, $U$, in the triplet pairing and
formation of other phases, we varied the $U$
values from ($U= 0.0$ to $3.0$), for fixed values of
attractive interaction $V_a=1.8$. 
 As shown in Fig.5(a) and Fig.5(b),
initially for lower values of $U$, the
correlation function, $C(r)$, shows nearly long range order,
while $P(r)$ decays exponentially, indicating CDW phase in
the system. On the other hand, for $U \gtrsim 1.1$, the
 correlation function, $P(r)$, shows algebraic decay
behaviour, displaying $TSF$ phase in the system.
To find out the phase boundary between the $CDW$ and 
$TSF$ phase, we have done finite size scaling of order-parameter
$O_p$. As shown in Fig.6(a), $O_p$ takes finite non-zero values
for $U=1.1\pm0.05$, indicate transition from $CDW$ phase to $TSF$ phase. 

As shown in Fig.5(c) and Fig.5(d),
with increase in $U$, initially $P(r)$ shows power law behaviour,
while $S(r)$ decays exponentially. On the otherhand, 
for large values of $U$, $S(r)$ shows power law behaviour,
while $P(r)$ decays exponentially. 
For moderate values of $U$, TSF and SDW phases compete
with each other. To find the phase boundary between 
$TSF$ and $SDW$ phase, we have done finite size scaling of
exponent of correlation function $S(r)$, as discussed in
previous section. Fig.6(b), shows transition from $TSF$ to
$SDW$ phase at $U=1.9\pm0.06$, as exponent of $S(r)$, takes the value $K=1$.
As shown in schematic of Fig.6(d),
we find CDW-phase for $U\lesssim 1.1$, TSF phase
for $1.1 \lesssim U \lesssim1.9$ and SDW phase for $U\gtrsim 1.9$,
for a fixed value of attractive interaction, $V_a=1.8$

\begin{figure}[h]
\rotatebox{0}{\includegraphics*[width=\linewidth]{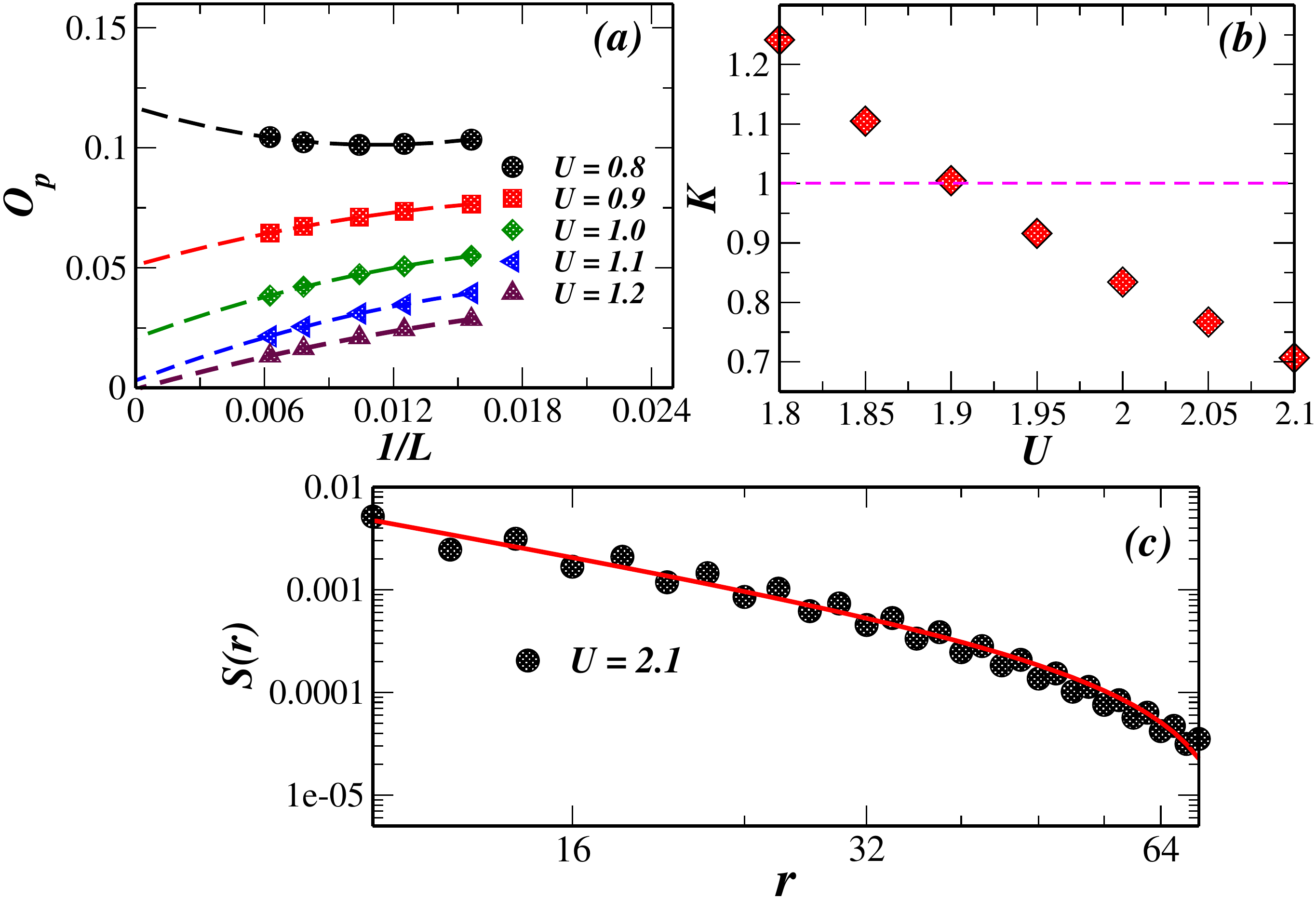}}
\rotatebox{0}{\includegraphics*[width=\linewidth]{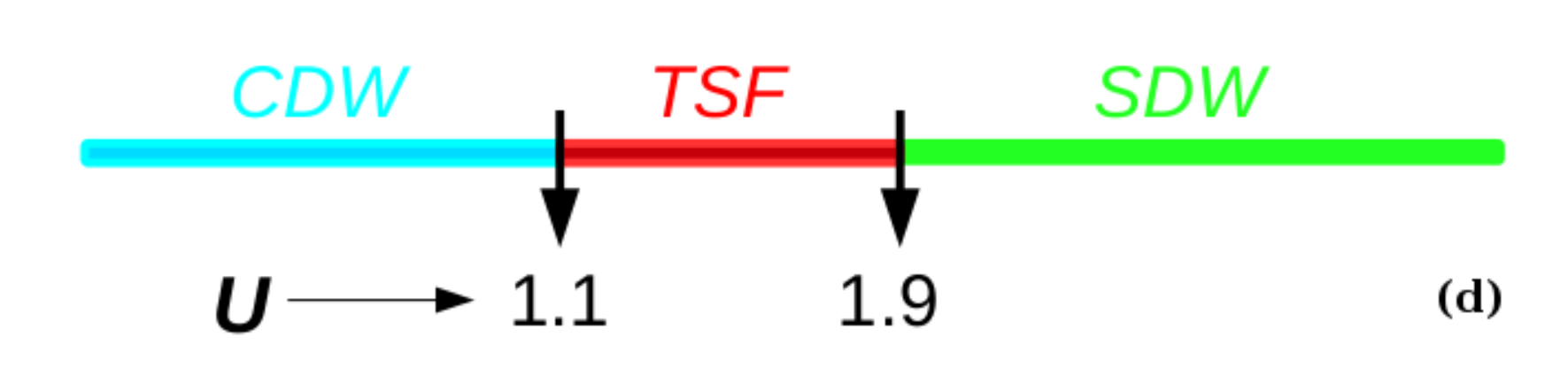}}
\caption{Finite-size scaling of (a) order parameter $O_p$ (b) exponent $K$ of the correlation function
$S(r)$, at $V_a=1.8$ and different values of $U$. 
(c)Power law fitting of $S(r)$ at $U=2.2$, on a log-log scale, for $L=128$.
(d) phase diagram for fixed value of $V_a=1.8$ with varying $U$.}
\end{figure}

\subsection{Effect of inter chain hopping}
\begin{figure}[h]
\rotatebox{0}{\includegraphics*[width=\linewidth]{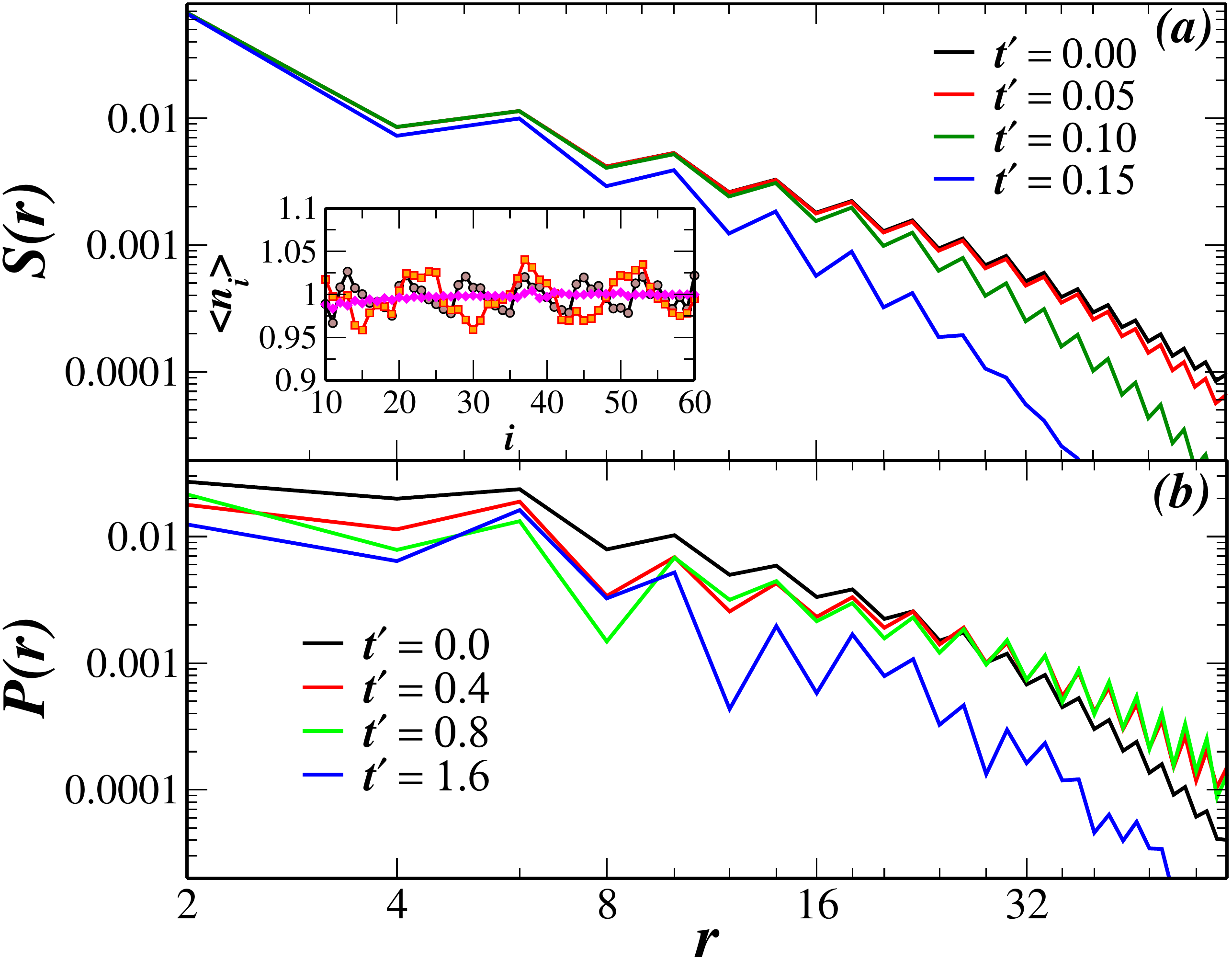}}
\caption{
(a) Plot of correlation function $S(r)$, (b) Plot of correlation function $P(r)$, 
as a function of $r$, at $U = 2$, $V_a=1.6$ and varying $t'$(on a log-log scale).
In the inset, charge density $\langle n_i \rangle$ is shown for $t'=1.2$ (circle), 
$1.6$ (square) and $2.4$ (diamond)} 
\end{figure}
Here, we study the effect of inter chain hopping, $t'$ on the 
triangular ladder. We find that, as the interchain hopping is
turned on, the SDW phase becomes unstable and disappears 
quickly with increase in $t'$. On the other hand, TSF phase 
becomes prominent with nonzero $t'$ values, however, as the
$t'$ becomes larger, the prominence decreases. The spin triplet 
pairs formed due to $V_a$ term along the rung, gets higher
stability with introduction of $t'$, as it promotes the 
antiferromagnetic exchange between the electrons on the rungs.
This results in increase in pair-correlation, $P(r)$. 
Interestingly, for large values of attractive interaction, 
$V_a$, when the system is in the CDW-phase, it gets hardly
affected by inter chain hopping term, as the charge
ordered state arrests the effective hopping between the chains. 
However, close to the phase boundary between TSF and CDW phases, when
the system is near the CDW phase boundary, for finite values of $t'$, 
system can again make transition to the TSF phase.

Now, using DMRG, we demonstrate the effect of $t'$ by
considering two values of $V_a$, $1.6$ and $2.8$, and
for a fixed value of $U=2$. These $V_a$ values correspond 
to SDW and CDW phases respectively, without any inter chain 
hopping term, $t'$. As we turn on $t'$, we look at the variation
in $SDW$ and $CDW$ phases. As shown in Fig.7(a), for $V_a=1.6$, 
the spin-spin correlation function, $S(r)$, 
starts decaying exponentially for $t'\gtrsim 0.1$ (Fig.7(a)), 
whereas, the pair correlation function, $P(r)$, 
initially increases with $t'$, for even small values of it. It clearly shows that the system 
makes transition from SDW phase to TSF phase in presence 
of interchain hopping $t'$. On the other hand, as we increase
the $t'$ value, for larger values of $t'$ ($t'\sim t $), the 
pair correlation function, $P(r)$, starts decreasing 
(Fig.7(b)). Interestingly, there the system show 
 a density profile, $\langle n(i)\rangle$, which is 
oscillatory in nature (as shown in the inset of Fig.7(a), for $t'=1.2$ and $1.6$).
In fact, at very large values of $t'$ ($t'\gtrsim 2.0$), 
the system enters into a metallic phase, where the density becomes 
homogeneous and takes values around one (see inset of Fig.7(a), for $t'=2.4$ ).
\begin{figure}[h]
\rotatebox{0}{\includegraphics*[width=\linewidth]{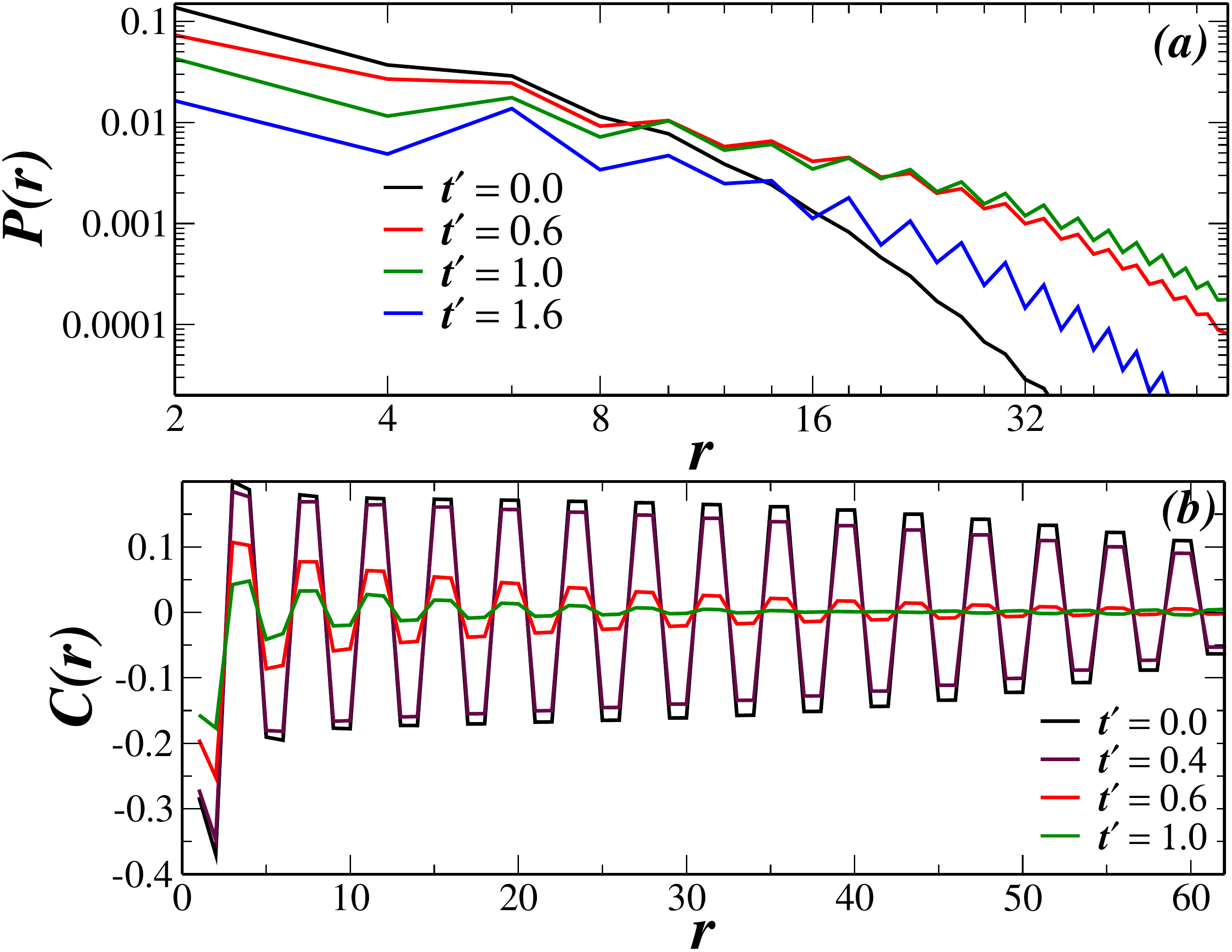}}
\caption{(a) Plot of correlation function $P(r)$ as a function of $r$ (on a log-log scale),
 (b) Plot of correlation function $C(r)$, as a function of $r$,
 at $U = 2.0$, $V_a=2.8$ with different values of $t'$.} 
\end{figure}

We find that the CDW-phase is quite robust against the 
interchain hopping term, $t'$. As shown in Fig.8(a), 
the charge-charge correlation function, 
$C(r)$, shows nearly long range order for $t'\lesssim 0.5$, 
On the other hand, as shown in Fig.8(b), the pair 
correlation function, $P(r)$, decays exponentially for lower values of $t'\lesssim0.5$,
while shows powerlaw behaviour for $t'>0.5$.
Such behavior of the correlation functions indicate a 
phase transition from CDW-phase to TSF phase for 
$t'\simeq  0.55 \pm 0.05$. 
For moderate values of $t'$, the $P(r)$ 
shows power law behaviour, while for larger values of $t'\gtrsim 1.2$,
 it starts decaying exponentially and the system again enters into a density wave phase.
For large values of $t'$ ($t' \sim 2.0$), the density wave phase enters into a metalic phase. 
For $V_a \gtrsim 3$, the CDW-phase is quite stable and it requires a really 
large values of $t'$ to destroy the CDW-phase.

\subsection{Effect of Intersite Repulsive Interactions}
\begin{figure}[h]
\rotatebox{0}{\includegraphics*[width=\linewidth]{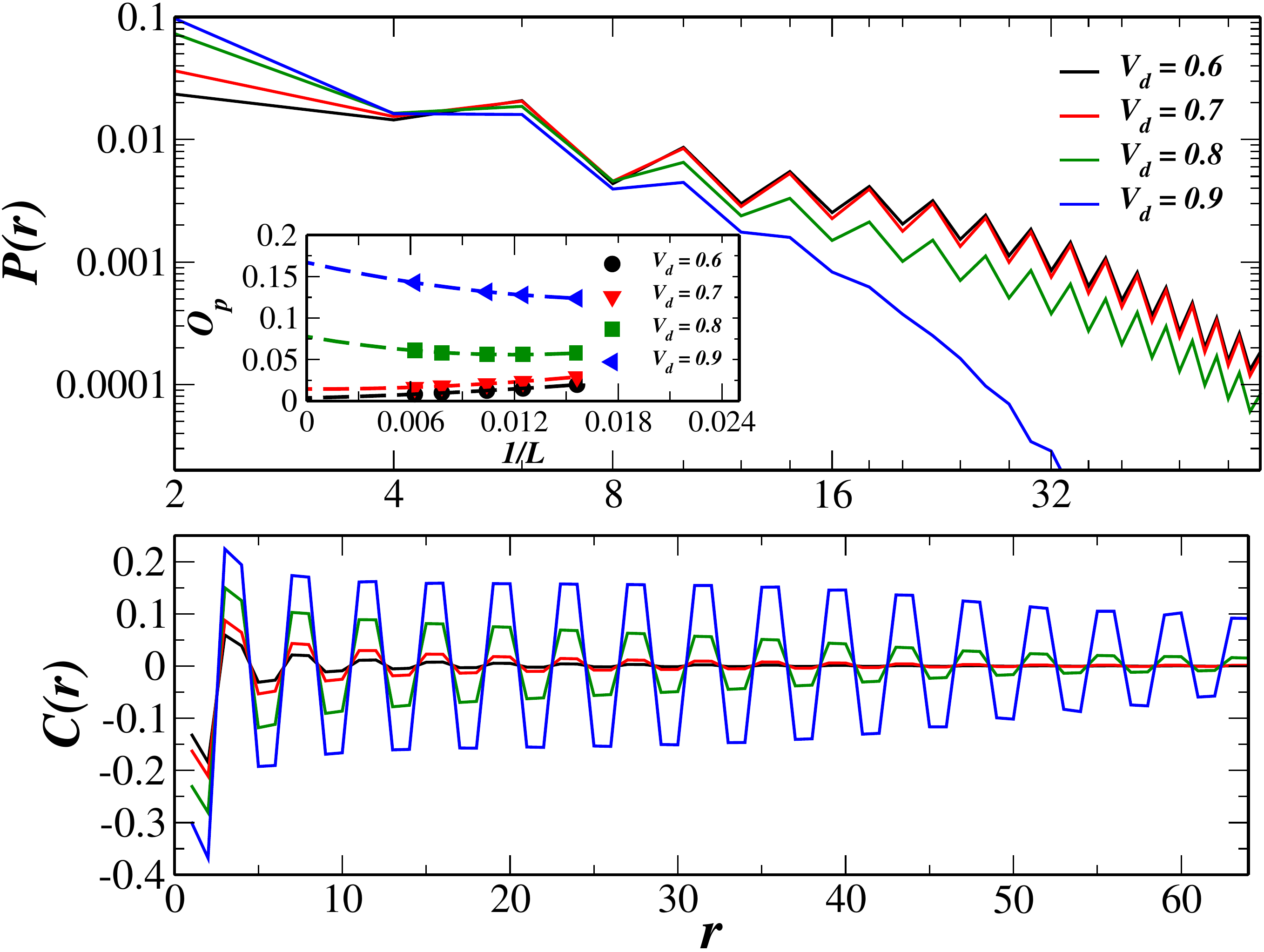}}
\caption{(a) Plot of correlation function $P(r)$, as a function of $r$, at $U = 2.0$,
$t'=0.4$, $V_a=1.8$ and different values of $V_d$. 
Inset shows finite size scaling of $O_p$ with $1/L$.
(b) Plot of correlation function $C(r)$, with distance $r$ at $U = 2.0$,
$t'=0.4$ ,$V_a=1.8$ and different values of $V_d$.}
\end{figure}

When the dipolar fermions are aligned along the rungs of 
the triangle, repulsive interactions can be generated along 
each chain direction $(V_r)$ as well as along the diagonal $(V_d)$ of the 
triangular ladder (as shown in schematic Fig.1).
For demonstrating the effect of repulsive interactions, $V_r$ and $V_d$, 
we chose interaction parameters $U=2$, $t'=0.4$, $V_a=1.8$ and vary the 
intersite repulsive parameters, $V_r$ and $V_d$.
 As discussed in the previous section, in absence of repulsive intersite
interactions, for these parameter values, the system remains in the TSF phase.
 On the other hand, with increase of intersite repulsive interactions, 
the fermions try to avoid each other and form a 
CDW state with structure, like $|2,0,0,2..\rangle$. 

In Fig.9, we have shown the effect of intersite repulsive interaction $V_d$, on the $TSF$ phase
keeping $V_r=0$. As shown in the Fig.9, for lower vales of $V_d < 0.8$, 
correlation function $P(r)$, shows power law behaviour (Fig.9(a)). For larger values of
$V_d$, correlation function $C(r)$, shows nearly long range behaviour (Fig.9(b)). 
To find the phase boundary between $TSF$ and $CDW$, we have done finite size scaling of $O_p$.
As shown in inset of Fig.9(a), $O_p$ takes small finite value for $V_d\sim0.7$.
In some cases, due to slow nature of transition and finite size effect, $O_p$ can take
very small non-zero values. So from plot of correlation function, $C(r)$ (Fig.9(b)) and
finite size scaling of $O_p$, we have estimated the transition from $TSF$ to $CDW$ phase at $V_d=0.75\pm0.06$.

\begin{figure}[h]
\rotatebox{0}{\includegraphics*[width=\linewidth]{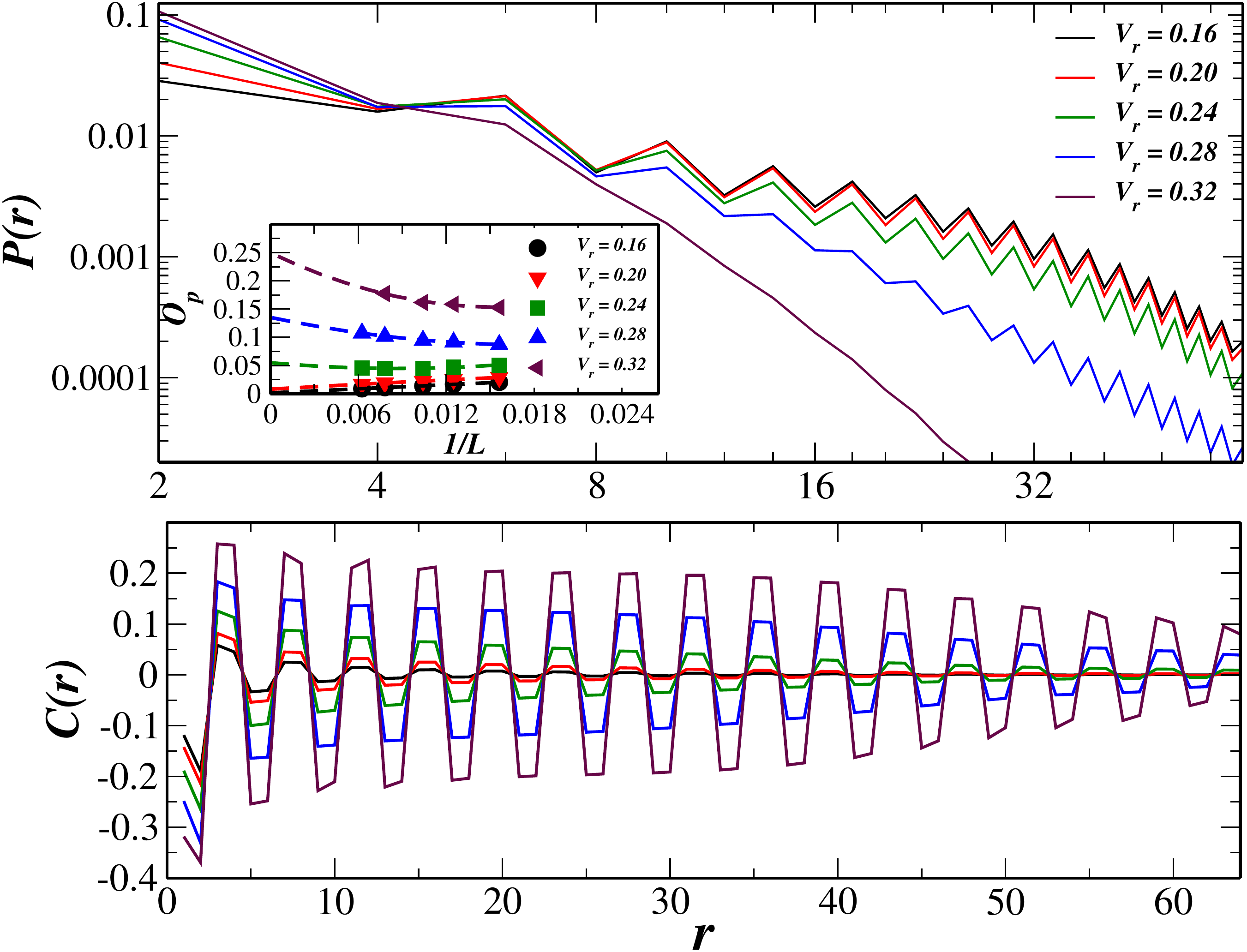}}
\caption{Plot of correlation function (a) $P(r)$, as a function of $r$, 
(b) $C(r)$, as a function of $r$ at $U = 2.0$, $t'=0.4$, $V_a=1.8$, $V_d=0.3$
 and different values of $V_r$. Inset shwos, finite size scaling of $O_p$ with $1/L$.}
\label{fig3}
\end{figure}

In the presence of attractive interaction, $V_a$, along 
the rungs of the triangles, the fermions in each of the 
chain become correlated with each other. 
We also found that in presence of $V_d$, 
small values of repulsive interaction $V_r$ is enough 
to produce a CDW-phase \cite{bpandey}. As shown in Fig.10(a),
 the pair correlation function, $P(r)$, shows power law behaviour up to
$V_r\sim 0.24$, while for larger values of $V_r$, it decays exponentially.
On the other hand, the charge charge 
correlation function $C(r)$, shows nearly long range 
behaviour for $V_r \gtrsim 0.24$ (Fig.10(b)). 
To find the phase boundary,
we have done finite size scaling of order-parameter, $O_p$.
As shown in the inset of Fig.10(a), $O_p$ takes finite value for $V_r=0.24\pm0.02$,
which clearly shows the phase transition from the $TSF$ phase to the $CDW$ phase at 
$V_r= 0.24\pm 0.02$.

\subsection{Effect of Three-body interaction}
\begin{figure}[h]
\rotatebox{0}{\includegraphics*[width=\linewidth]{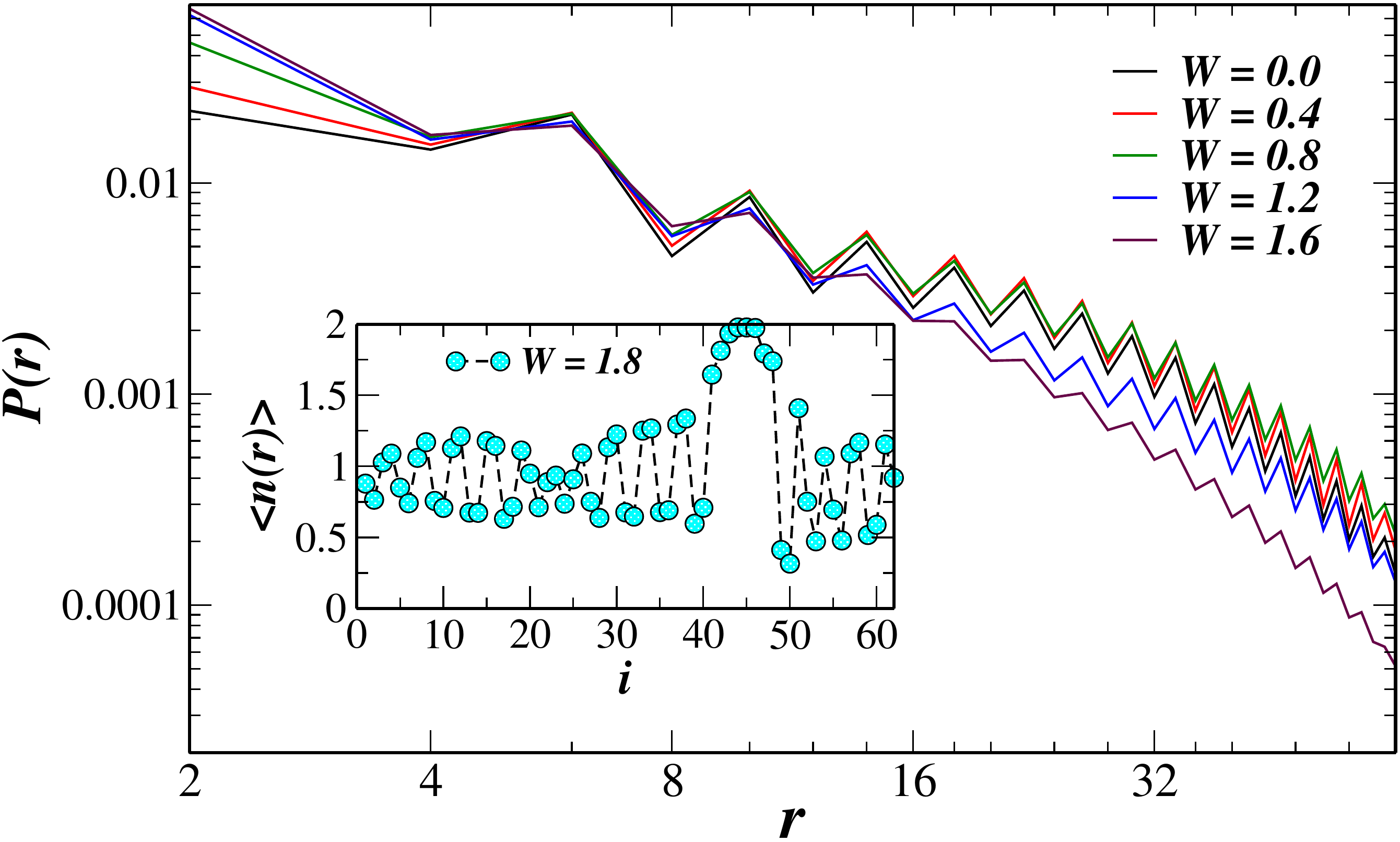}}
\caption{Plot of correlation function, $P(r)$, as a function of $r$, for interaction parameters, 
$U = 2.0$, $V_a=1.8$, $V_r=0.1$, $V_d=0.3$, $t'=0.4$ and different values of $W$. 
Inset shows, density profile of fermions $\langle n_i \rangle $,
with site index $i$, for $W=1.9$.} 
\label{fig11}
\end{figure}
Due to triangular geometry and dipolar interactions, 
an additional three-body interaction term may appear 
in each of the triangular plaquette, as suggested by others 
on similar grounds \cite{lbonnes,xuefeng}. 
Three body term can break the particle hole symmetry 
of the Hamiltonian. In optical lattices, the three body 
and two body interactions can be tuned independently\cite{hpb,kps}. 
Here, we demonstrate the consequences of attractive three body  
interaction\cite{swill,tsowi}, $W$, along with two body interactions and ask
whether the three body term can generate new phases or combine
several phases. To show the effect of three body interactions, 
we choose the system parameters, $U=2$, $V_a=1.8$, $V_d=0.3$, $V_r=0.1$ 
and $t'=0.4$ and varied the $W$. Without $W$ term, the system
exists in TSF phase for these parameters. As we turn on the 
attractive three body interaction, $W$, both $TSF$ and $CDW$ 
phases coexist and the system remains so up to moderate 
values of $W$. 

As shown in the Fig.11, triplet pair correlation function, $P(r)$ 
with increase in $W$, shows power law behavior, with slight changes in exponent.
 Additionally, with increase in $W$, a periodic modulation appeared in the 
charge charge correlation function, $C(r)$. To see
the appearance of $CDW$ order in the thermodynamic limit, we 
have done finite size scaling of order-parameter, $O_p$. 
As shown in inset of Fig.12, $O_p$ 
takes finite nonzero values for $W=0.6\pm0.1$.
Periodic modulation in density correlation, $C(r)$, and 
algebraic decay of $P(r)$,  
give signature of fermionic supersolid phase 
in the system for $0.6\lesssim W \lesssim1.7$, where both $CDW$ and $TSF$ phases coexist.
This supersolid phase is different from the supersolid phase formed due to coexistence of onsite 
pairing of fermions ($s$-wave superfluid), and charge density wave of the system. 
Here, fermions form pairs in spin-triplet sate ($p_z$-wave superfluid),
 which coexist with $CDW$ phase of the system.  
For large values of $W \gtrsim 1.7$, the system 
becomes unstable and thereby become phase separated.
In the phase separated state, density distribution is inhomogeneous,
while correlation function, $P(r)$ decay exponentially. Note that,
in the phase separated state, there is generally convergence problem, which
we found for $W \gtrsim 2.0$. In the inset of Fig.11, 
plot of charge density profile $\langle n_i \rangle$ 
has been shown for $W=1.8$, with site index $i$ (also see in Appendix Fig.16, 
the plot of $\langle n_i \rangle $, for different values of $W$). 
\begin{figure}[h]
\rotatebox{0}{\includegraphics*[width=\linewidth]{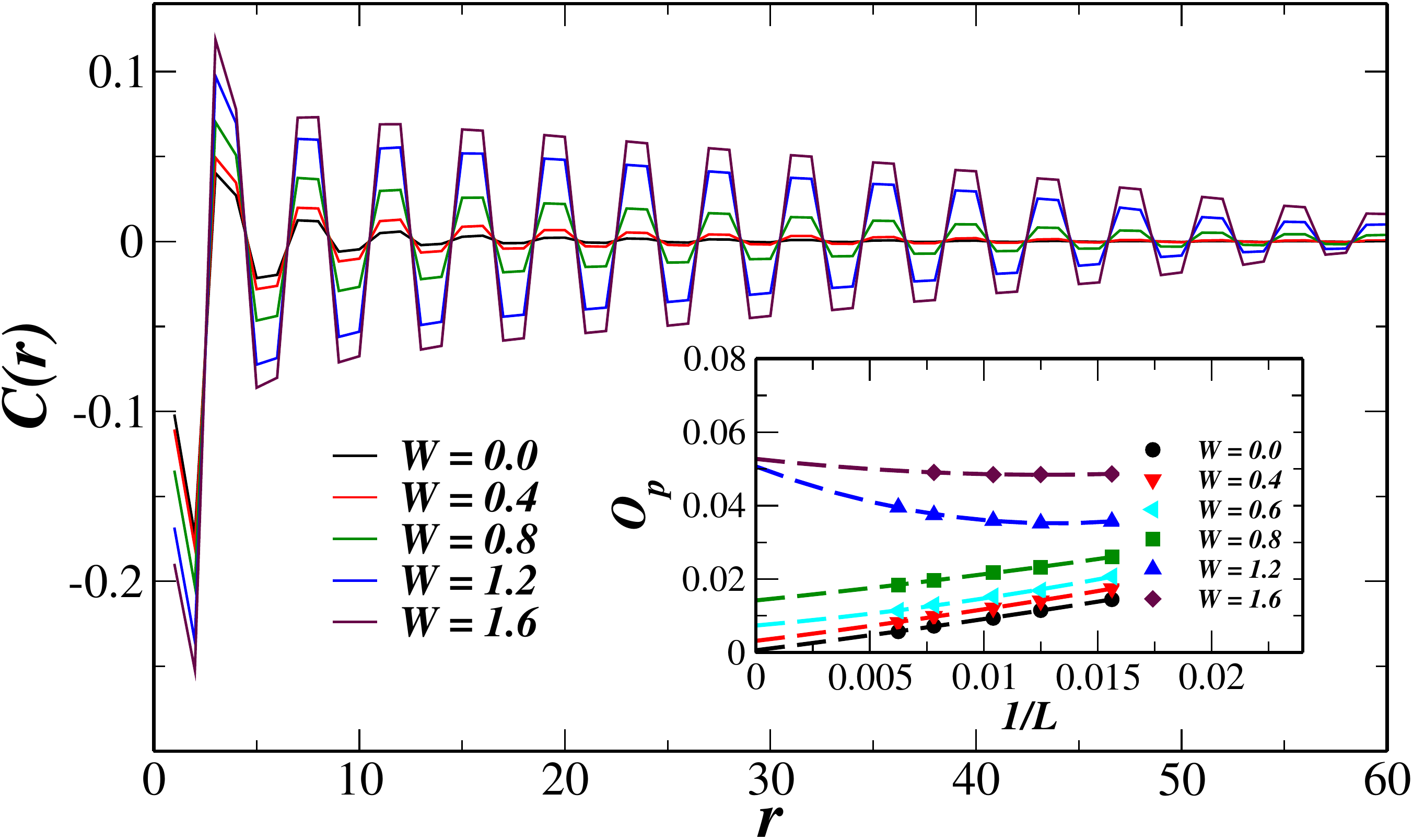}}
\caption{Plot of correlation function, $C(r)$, as a function of $r$, for interaction parameters, 
$U = 2.0$, $V_a=1.8$, $V_r=0.1$, $V_d=0.3$, $t'=0.4$ and different values of $W$. 
 Inset shows, finite size scaling of $O_p$ with $1/L$.} 
\end{figure}

\subsection{Effect of spin-dependent hopping}
\begin{figure}[h]
\rotatebox{0}{\includegraphics*[width=\linewidth]{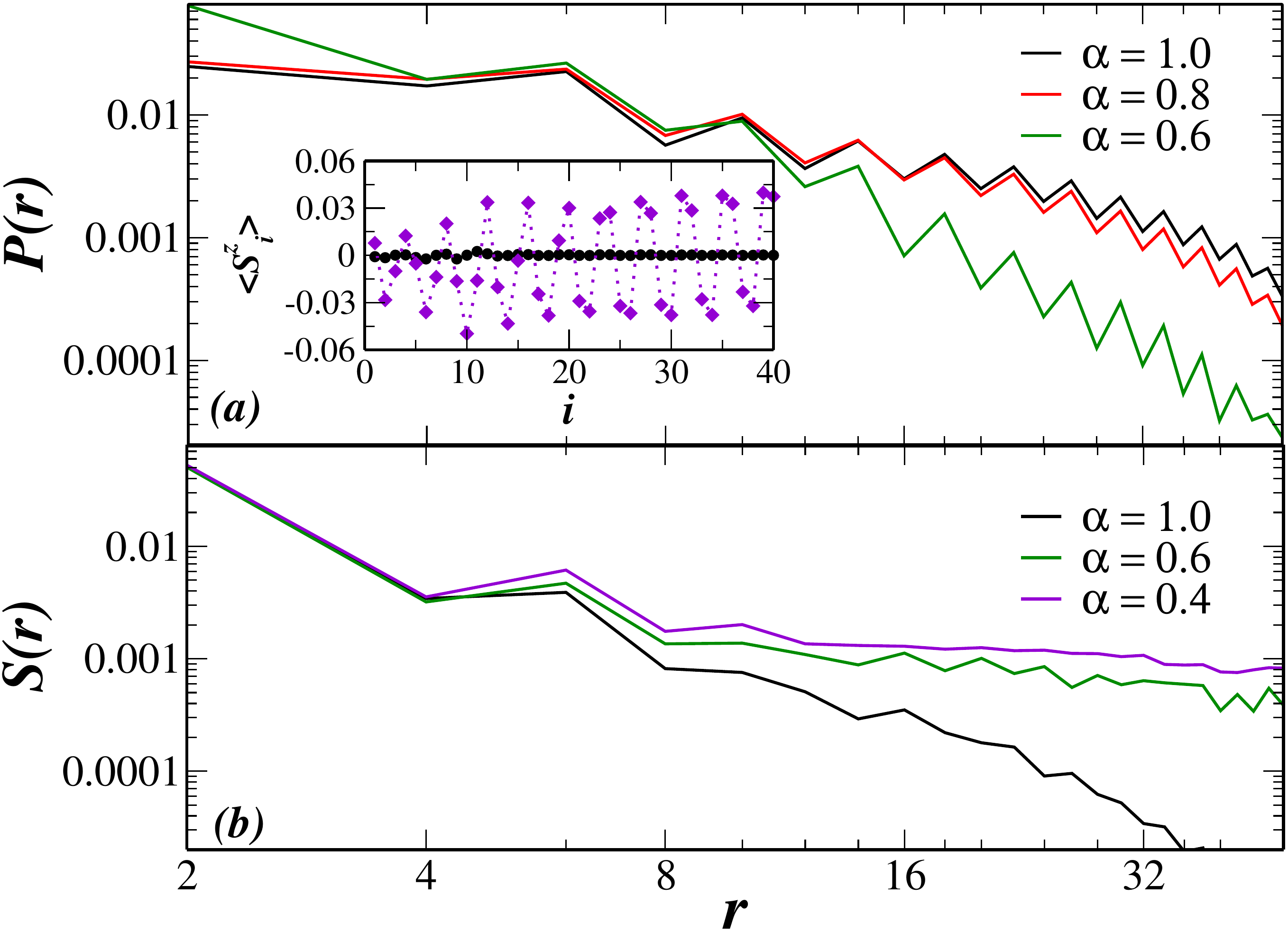}}
\caption{
 (a) Plot of correlation function $P(r)$, (b) Correlation function $S(r)$,
 as a function of $r$, at $U=2$, $V_a=2.0$, $V_r=0.1$, $V_d=0.2$ and $t'=0.4$, with varying $\alpha$.
Inset shows, plot of spin density $\langle S_i^z \rangle$ with site index $i$, 
for $\alpha=1$ (circle) and $\alpha=0.4$ (diamond).}
\end{figure}
In this section, we analyze the effect of spin dependent 
hopping on the TSF phase. We apply spin dependent hopping 
along the rungs of the triangle. We considered up-spin hopping
term to be more stronger than the down-spin hopping term
\cite{wvliu}. The corresponding change in hopping term in the 
Hamiltonian can be written as
\begin{equation*}  
H_{t_{\sigma}}=\sum_i \left(t'_{\uparrow} c^+_{i,\uparrow}c_{i+1,\uparrow}+h.c\right)
+\left(\alpha t'_{\downarrow} c^+_{i,\downarrow}c_{i+1,\downarrow}+h.c\right)
\end{equation*}
\noindent where, $\alpha$ is an anisotropic term 
($\alpha=1$ make the Hamiltonian same as Eq.1). 
Spin dependent hopping term breaks the spin rotational symmetry,
$SU(2)$ and also the time reversal symmetry of the 
Hamiltonian\cite{macazilla,yehua}. As the $SU(2)$ 
symmetry is broken, ground state is no more in $s_{tot}^z=0$ sector, while the number sector is 
still fixed. In such a situation, we have checked our DMRG results
 with those from exact diagonalization results with the same setup for smaller 
system sizes. As the results compare fairly well, we have set up DMRG calculations with 
fixed number of particles without considering $s_{tot}^z$ quntum number. Since the matrix 
dimension in each of the DMRG iteration increases quite considerably ($\sim10^6$), 
we have carried out DMRG calculations with max=$450$ and for system length, $L= 96$.
We have verified the results for $\alpha =1$ by running DMRG calculations with $s_{tot}^z=0$
and without considering $s_{tot}^z$ quntum number upto $L=96$ with max$=450$ and found the results 
compare quite well. We thus have carried out DMRG calculations with 
the paramaters, $U=2$, $V_a=2$, $V_r=0.1$, $V_d=0.2$, $t'=0.4$ with
 varying $\alpha$ values. Note that, for these parameter values with $\alpha = 1$, the 
system is known to be in TSF phase (see Fig.13(a)). 

 With spin-dependent hopping, we find that the TSF phase is 
suppressed, while the SDW phase starts dominating.
As shown in Fig.13(a), the pair correlation function, $P(r)$, 
decays algebraically for $\alpha \gtrsim 0.6 \pm 0.1$, 
showing clearly that the TSF phase is sustained by 
spin dependent hopping, while for $\alpha \lesssim 0.6\pm 0.1$,
the pair correlation decays exponentially. With lower 
values of $\alpha$, spin-spin correlation function, $S(r)$, 
has nearly quasi long range order for $\alpha \lesssim 0.6 
\pm 0.1$ (Fig.13(b)). In the inset of Fig.13(a), we show 
spin density profile, $\langle s^z_i\rangle$ with site index $i$. 
As can be seen, the $\langle s^z_i\rangle$ takes finite values for 
$\alpha=0.4$, however, it vanishes for $\alpha = 1.0$. 
For lower values of $\alpha$, the down-spin becomes reluctant 
to hop between legs of triangle, thus promoting SDW phase while 
suppressing TSF phase in the ladder system.

\section{Conclusion}

In summary, we have investigated the SDW, TSF and CDW phases of
dipolar fermions, at half filling, on a triangular ladder.
In presence of moderate values of repulsive onsite interaction and
attractive intersite interactions, the fermions form
exotic spin triplet superfluid phase. In presence of
intersite attractive interactions, and onsite repulsive interaction,
a charge density wave phase is found even without any intersite
repulsive interactions. We have demonstrated the stability of
spin triplet phase, by introducing inter leg hopping, which effectively
enhances the spin triplet superfluid phase region by replacing the
spin density wave phase.
In presence of repulsive interactions,
we show transition between TSF phase and a CDW phase. We
also have looked at the effect of three body interactions on
the TSF and CDW phases. We find that the three body term
can introduce a fermionic supersolid phase,
where both TSF and CDW coexist. We strongly believe that our study, 
which unravel the rich physics of exotic phases of dipolar-fermionic systems
in ultra-cold systems would show inroads for further experiments.

\section{Acknowledgments}
B.P. thanks the UGC, Govt. of India for support through 
fellowship and S.K.P. acknowledges DST, Govt. of India 
for financial support.

\section{Appendix}

\begin{figure}[h]
\rotatebox{0}{\includegraphics*[width=\linewidth]{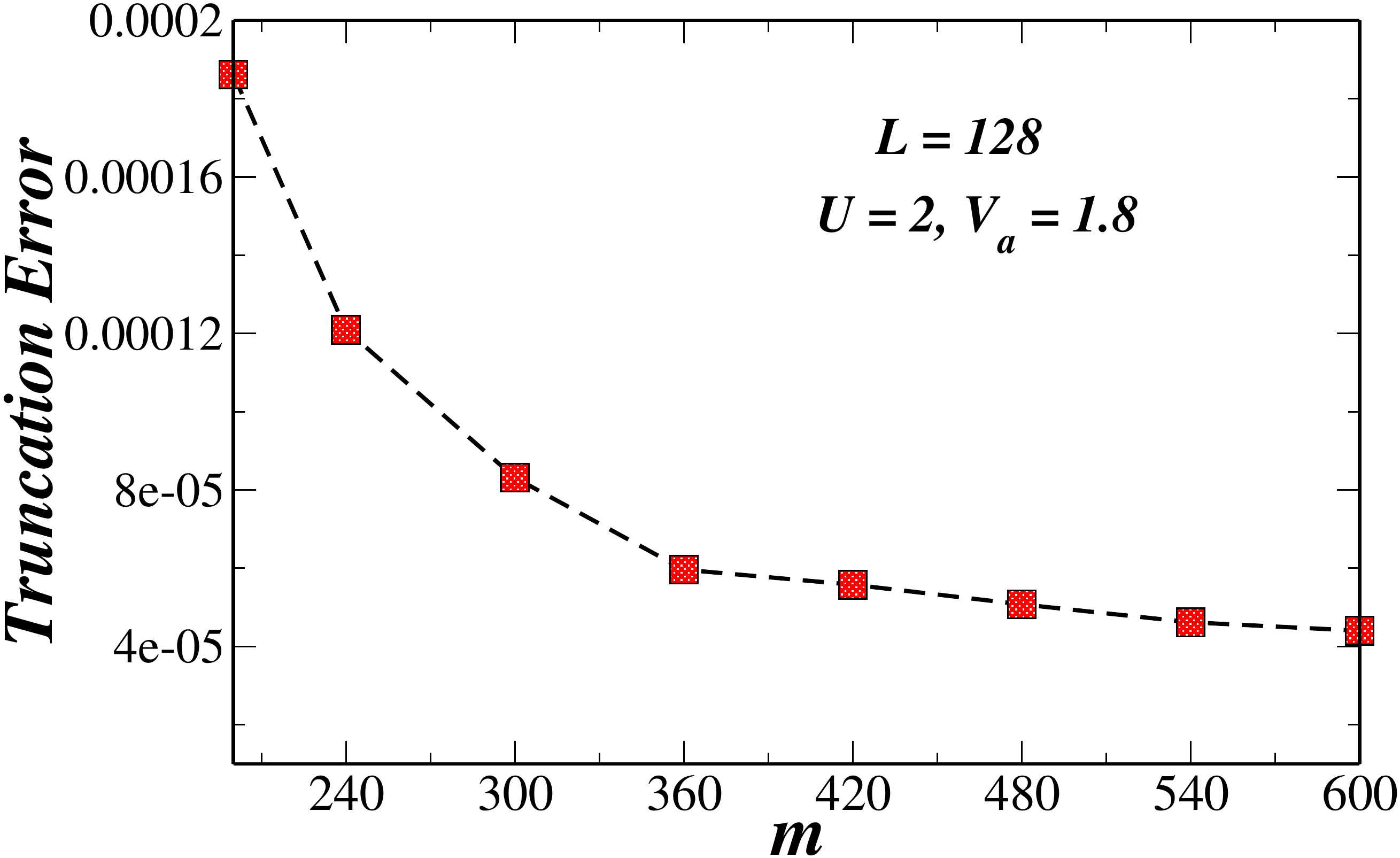}}
\caption{Plot of truncation error with max values m, for interaction parameters
 $U=2$, $V_a=1.8$  {\bf{ (other parameters are kept zero)}}.
}
\end{figure}
To check the accuracy of our DMRG calculations, we have calculated truncation error of the system.
In DMRG, the effective  basis is truncated by keeping the $m$ largest eigenvectors of the reduced 
density matrix corresponding to the $m$ largest eigenvalues. The error caused by the
truncation can be measured by calculating $e=1-\sum_i \rho_i$, 
where $\rho_i$ is the eigenvalues corresponding to the reduced density matrix.
Fig.14 shows plot of truncation error with max values $m$, for system size $L=128$ and 
{\bf{for interaction parameters values $U=2$, $V_a=1.8$, keeping all the other parameters,
 $t'$, $V_r$, $V_d$ and $W$, as zero}}. With increase in max value $m> 420$,
truncation error changes very slowly.

\begin{figure}[h]
\rotatebox{0}{\includegraphics*[width=\linewidth]{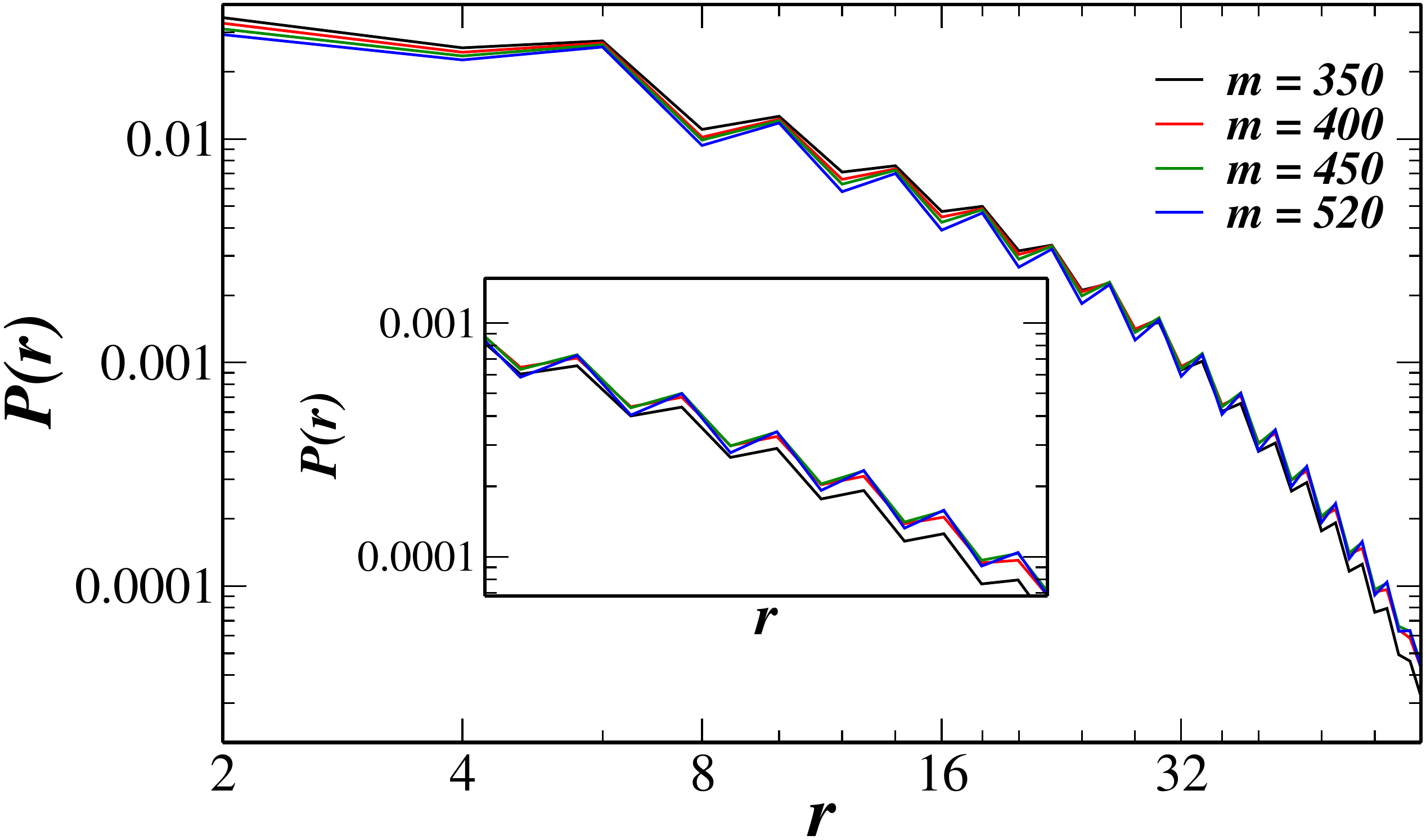}}
\caption{ Plot of correlation function, $P(r)$, as a function of $r$, at $U=2$, $V_a=1.8${\bf{ (other
parameters are kept zero)}}, with different max values.}
\end{figure}

To check the behaviour of correlation function, $P(r)$, with max values,
we have calculated the correlation function with different max values (Fig.15).
As shown in inset of Fig.15, $P(r)$ almost overlaps for $m=450$ and $m=520$.
This proves that $m$ value of $450$, is large enough to obtain accurate correlation 
function, $P(r)$. 

We have used open boundary condition for our calculations in $DMRG$.
To remove the edge effects, we have computed correlation functions from
central site to one side of the triangular ladder.
In case of correlation functions $S(r)$ and $P(r)$, we found that with increase in
distance $r$, rapid fluctuations appeared in correlation functions.
\begin{figure}[h]
\rotatebox{0}{\includegraphics*[width=\linewidth]{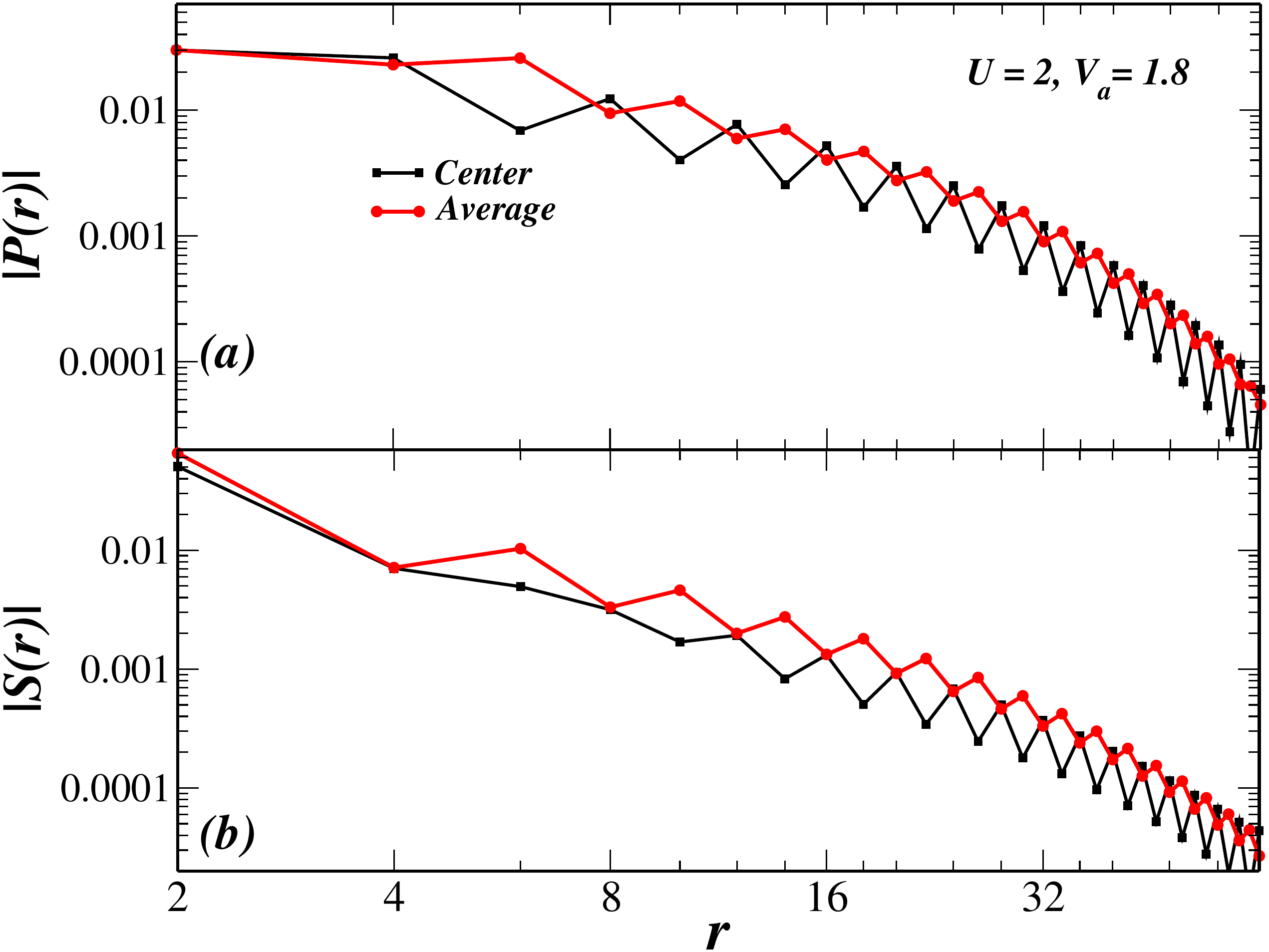}}
\caption{Plot of correlation functions for interaction parameters
 {\bf{$U=2$, $V_a=1.8$ (other parameters are kept zero)}} 
(a)$S(r)$ and  (b)$P(r)$, in two different way, one from center of
the lattice (square) and second by taking average (circle) for system size $L=128$.}
\end{figure}
As shown in Fig.16, to smoothen these fluctuations, 
we have calculated average correlation function,
$S(r)= 1/N(r)\sum_r \left|\langle s_i^zs_{i+r}^z\rangle \right|$,
where we took the sum over the correlations, which are separated
by the same distance $r$ from the sites $i$ from the other side of the ladder.
This is then divided by the number, $N(r)$, of such same distance correlations.
While averaging, {\bf{we excluded lattice sites within distance  
$L/4$ from both the end of the ladder (of system size, $L$)}}. 
We have calculated the average correlation function, 
$P(r)= 1/N(r)\sum_r |\langle \Delta_l^+ \Delta_{l+r}\rangle|$,
by summing over the correlations, which are separated by 
same distance $r$ and divide the sum by $N(r)$.

\begin{figure}[h]
\rotatebox{0}{\includegraphics*[width=\linewidth]{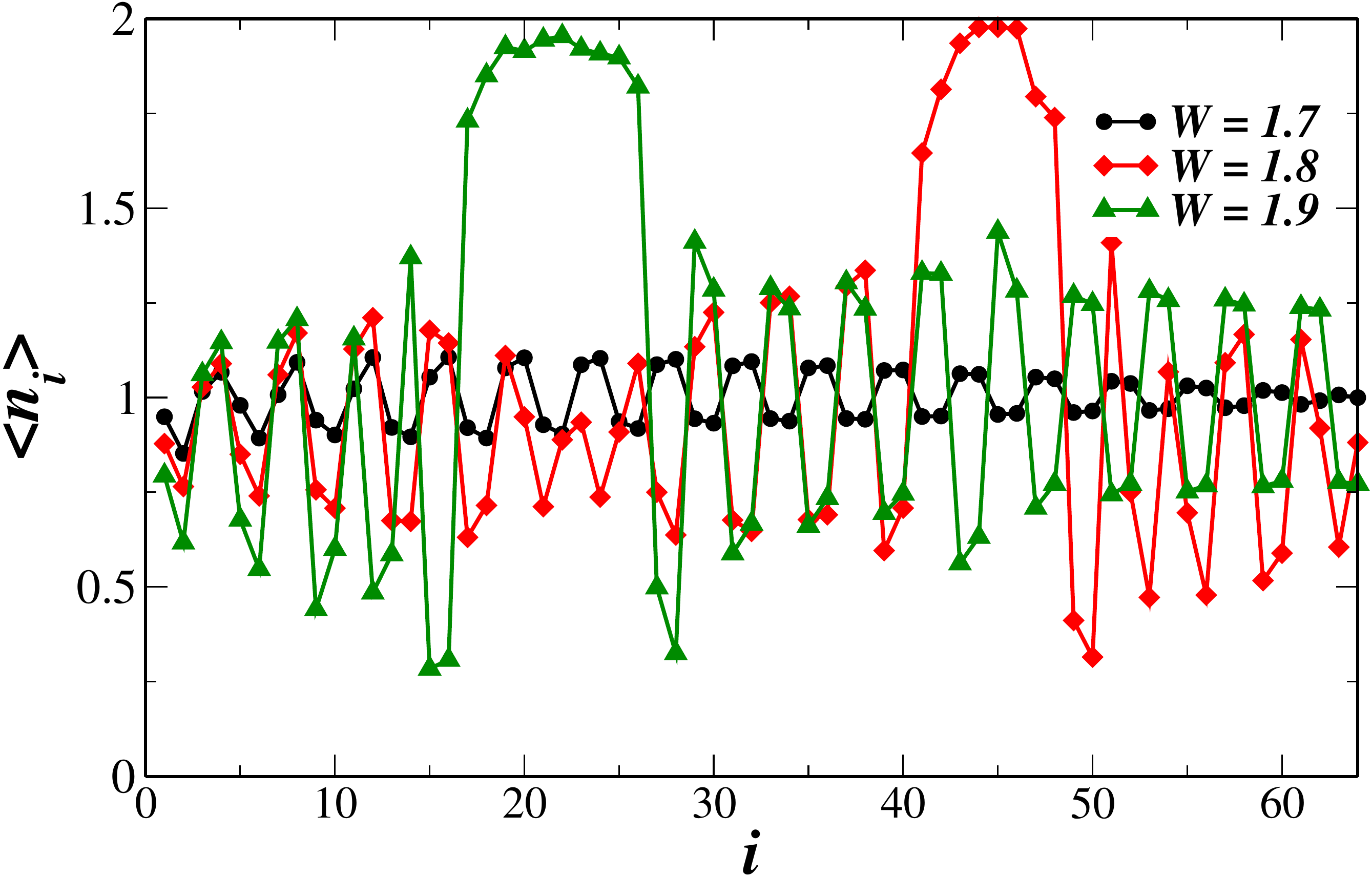}}
\caption{Plot of charge density $\langle n_i \rangle $, 
for interaction parameters, $U=2.0$, $V_a=1.8$,
$V_r=0.1$, $V_d=0.3$, $t'=0.4$ and different values of $W$. 
}
\end{figure}

As discused in section III.E, for large values of $W\gtrsim 1.7$, 
the system enters into a phase separated state. In Fig.17, 
the plot of charge density profile $\langle n_i \rangle$ has been shown, for 
interaction parameters, $U=2.0$, $V_a=1.8$, $V_r=0.1$, $V_d=0.3$, $t'=0.4$ 
and three diffrent values of $W$. For $W=1.7$, the system shows a periodic density modulation,
while for $W=1.8$, inhomogenious feature appears in the density profile.
Interestingly, for $W=1.8$, the $\langle n_i \rangle$ takes the maximum possible values 
($\sim 2$) near the center of the ladder, and for $W=1.9$, it shifts to one side of the ladder.
For $W\gtrsim 2.0$, we found convergence problem.

\end{document}